\documentclass[fleqn,10pt]{wlscirep}
\usepackage[utf8]{inputenc}
\usepackage[T1]{fontenc}
\usepackage{subcaption}
\usepackage{caption}

\title{Temporally Modulated One-Dimensional Leaky-Wave Holograms}

\author[1]{Amrollah Amini}
\author[2,*]{Homayoon Oraizi}
\affil[1]{Iran University of Science and Technology, School of Electrical Engineering, Tehran, 1684613114, Iran}

\affil[*]{h\_oraizi@iust.ac.ir}

\affil[+]{these authors contributed equally to this work}

\begin{abstract}
Spatio-temporally modulated impedance surfaces can be good candidates for generation of radiating waves with arbitrary eigenstates by breaking momentum and energy conservations.
Here, we present a theoretical framework based on the holographic technique and generalized Floquet-wave expansion  to analyze spatio-temporally modulated impedance surfaces.
The holographic technique estimates the required  impedance distribution to achieve the desired momentum.
Injecting temporal modulation deviates the eigenvalues and changes the radiation frequency.
Using the proposed analytical  model, the eigenvalues can be calculated accurately in the presence of space and time modulations. Consequently, it is possible to predict the propagation mechanism of bounded and radiation states.
It has been shown that, imposition of temporal modulation causes  the Doppler-shift effect and nonreciprocal responses in the hologram. 
By plotting the antenna dispersion diagram, and observing the asymmetric displacement of dispersion curve due to temporal modulation, the system nonreciprocity can be verified.  The beam scanning properties of these structures have also been investigated.

\end{abstract}
\begin{document}

\flushbottom
\maketitle
\thispagestyle{empty}

%\noindent Please note: Abbreviations should be introduced at the first mention in the main text – no abbreviations lists. Suggested structure of main text (not enforced) is provided below.

\section*{Introduction}
Recently, planar nonreciprocal devices without the use of bulky ferromagnetic materials have received much attention \cite{estep2014, sounas2017} due to their subwavelength thicknesses and low cost manufacturing processes. A suitable solution for implementing such devices is to exploit spatio-temporally modulated structures.
This method can cover a wide range of frequencies from microwaves \cite{taravati2020_mm} to optical \cite{shaltout2019, shaltout2019_2} regimes.
The use of spatio-temporally modulated structures is not limited to electromagnetic devices, but can also be applied to manipulate acoustic waves \cite{fakheri2021}.
A wide variety of potential applications can be envisaged for space-time structures, such as isolators \cite{yu2009}, frequency translators \cite{wu2020}, circulators \cite{sounas2013}, and nonreciprocal phase shifters \cite{wang2020}.
In the context of antenna engineering several works have also been proposed to realize nonreciprocal radiators \cite{taravati2020_PRApplied, taravati2017,serrano2015}.
A mixer-duplexer leaky-wave antenna is presented in the literature using temporally modulated microstrip lines \cite{taravati2017}. An array of varactors incorporated in transmission lines was used to inject temporal modulation.
A silicon-compatible leaky-wave antenna has been designed based on graphene impedance surfaces \cite{serrano2015}. Nonreciprocity and Doppler-shift effects are obtained by applying temporal modulation through  gating pads located adjacent to graphene surfaces. These structures can be used for bio-sensing, imaging, and inter-chip communication applications \cite{serrano2015}.

An appropriate approach to implement spatio-temporally modulated structures is to use dynamic metasurfaces, which have several advantages such as ultra-thin thicknesses, low cost and high flexibility. Metasurfaces  are mainly exploited in transmissive \cite{khorasaninejad2016, ee2016, khorasaninejad2017, zhu2014, cai2015, yang2021}, reflective \cite{yang2014, park2017, nayeri2015, yang2017_TAP}, or leaky-wave \cite{faenzi2019,moeini2019} modes. The main drawback of microwave metareflectors and transmitarrays is their protruded feeds  which make them unsuitable for low-profile integrated systems. This shortcoming can be resolved by using leaky-wave metasurfaces which are generally fed by embedded monopole structures.
Owing to the above-mentioned attractive features, various methods have been recently proposed for the design and analysis of leaky-wave metasurfaces.
The holographic technique \cite{sievenpiper2005, fong2010, li2014}, aperture field estimation (AFE) method \cite{teniou2017,minatti2015,amini2020}, Floquet-wave expansion model \cite{minatti2016_FO,amini2021} and the method of moments (MoM) \cite{ovejero2015,bodehou2019} are among the proposed methods of designing metasurface-based leaky-wave antennas.
The holographic and AFE techniques can be considered as synthesis methods for the estimation of arrangement of  meta-atoms  to achieve the desired wavefronts, whereas the MoM and Floquet-wave expansion are analysis frameworks for the determination of exact solutions of leaky-wave modes generated by the synthesized metasurfaces. Therefore, for a comprehensive study of leaky-wave metasurfaces, we need to combine the analysis and synthesis methods.

In this paper, the combination of holographic technique and generalized Floquet-wave expansion is proposed to analyze the temporally modulated one-dimensional leaky-wave metasurfaces. The scalar impedance boundary condition is used to model the proposed hologram. The generalized Floquet-wave expansion was first proposed by Cassedy \cite{cassedy1965} to attain the mode characteristics of temporally modulated surface waveguides. In our work, this method is used for impedance surfaces  to explain the radiation mode and the leakage mechanism in the presence of space-time modulation.
As an example, using the holographic technique, the surface impedance is designed to radiate in the direction of $30^\circ$ at 18 GHz. The effect of temporal injection is studied and the radiation characteristics of antenna in the transmission and reception modes have been investigated.

\section*{Spatio-temporally modulated impedance surfaces}
\begin{figure}
	\centering
	\includegraphics[width = 0.55\textwidth]{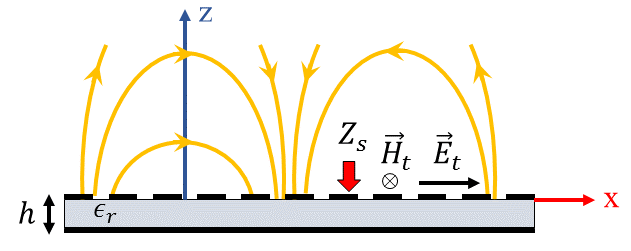}
	\caption{Surface impedance boundary condition implemented on a grounded dielectric slab. The propagation direction is assumed to be along the x-axis.}
	\label{fig:Zs_IBC}
\end{figure}

A convenient way to implement holographic metasurfaces in the microwave regime is to modulate impedance surfaces. For an impedance surface located at $z = 0$ (see Fig. \ref{fig:Zs_IBC}), the tangential components of electric ($\vec{E}_t$) and magnetic ($\vec{H}_t$) fields are related through the impedance boundary condition as follows \cite{tretyakov2003}:
\begin{equation}
	\vec{E}_t(x, y) = \underline{\underline{Z}}_s(x, y) \cdot \hat{z} \times \vec{H}_t(x, y)
\end{equation}
where $\underline{\underline{Z}}_s$ is the tensorial impedance indicating anisotropic boundary conditions and $\hat{z}$ denotes the unit vector along the z-axis.  For scalar (isotropic) impedance surfaces, the boundary condition is simplified as:
\begin{equation}
	\vec{E}_t(x, y) = Z_s(x, y) \hat{z} \times \vec{H}_t(x, y)
\end{equation} 
Figure \ref{fig:Zs_IBC} shows the conceptual representation of an impedance boundary condition. For spatially modulated surfaces, dielectric slabs covered by periodic subwavelength patches can be used for  the realization of impedance boundary conditions. Generally, impedance surfaces support both TM and TE surface modes. 
Note that for the TM (TE) mode the surface wave has transverse magnetic (electric) field components in both propagation and $\hat{z}$ directions.
In  \cite{oliner1959} it has been shown that if the surface impedance is modulated sinusoidally in the propagation direction, the power will leak from the surface at a certain angle which is proportional to the impedance periodicity. In this case, the surface impedance acts as a leaky-wave antenna.
However, this structure has several characteristics such as linearity and reciprocity. These properties may be rectified by adding temporal variations to the impedance surface.
In spatio-temporally modulated surfaces, the impedance boundary condition changes simultaneously in  space and time domains. In the general case, the impedance boundary condition can be expressed as follows:
\begin{equation}
	Z_s(x, y; t) = \sum_n Z_n e^{-jnKs(x, y; t)}
	\label{eq:Z_s_x_y_t}
\end{equation}
where $Ks$ is the modulation phase and represents the local periodic variations in the impedance boundary condition. 
The coefficient $Z_n$ can be a complex number and determines the variation depth of surface impedance.
For a monochromatic modulation, equation (\ref{eq:Z_s_x_y_t}) is simplified as:
\begin{equation}
	Z_s(x, y; t) = Z_0 + Z_{-1} e^{jKs(x, y; t)} + Z_{+1} e^{-jKs(x, y; t)}
\end{equation}
In this case we may define the modulation frequencies in space and time domains, which are indeed the spatial and temporal derivatives of $Ks(x, y; t)$, respectively:
\begin{equation}
	\vec{\beta}_p = \nabla_{x, y} Ks(x, y; t) = \frac{\partial}{\partial x} Ks(x, y; t) \hat{x} +  \frac{\partial}{\partial y} Ks(x, y; t) \hat{y}
\end{equation}
\begin{equation}
	\omega_p = -\frac{\partial}{\partial t} Ks(x, y; t)
\end{equation}

\begin{figure}
	\centering
	\includegraphics[width = 0.45\textwidth]{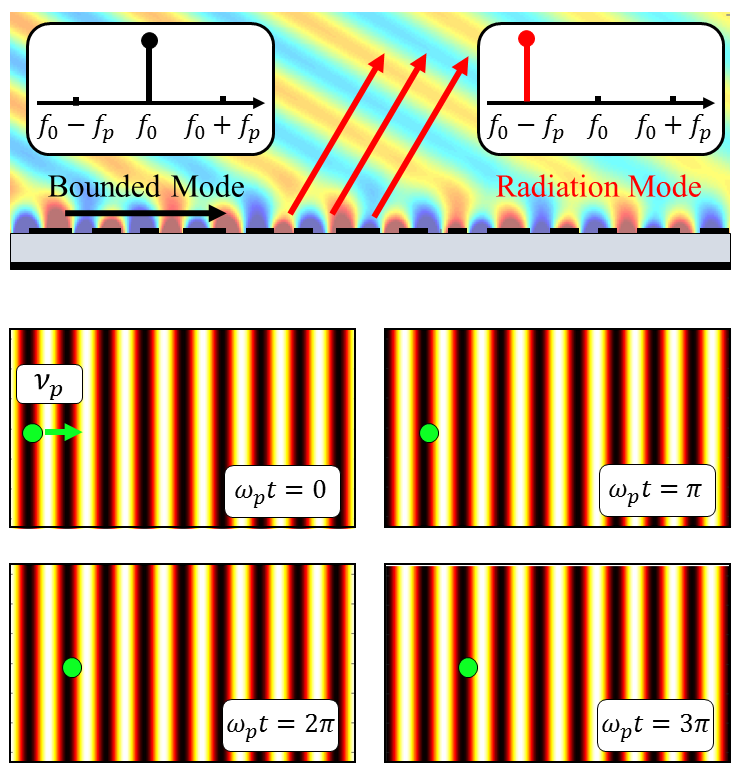}
	\caption{Scalar surface impedance patterns for different time slots. The circular indicator shows the progression of impedance with time.}
	\label{fig:Zs_TV_1D}
\end{figure}

The frequencies $\vec{\beta}_p$ and $\omega_p$ are actually space and time domain pumping frequencies, respectively, that cause power to be coupled from surface mode to higher order Floquet modes.
If the pumping frequencies are equal to zero, the power coupling will not take place and we will have a pure surface mode.
Note that, $Ks$ can be an arbitrary function of  (x, y) in the cartesian coordinate system.
In this paper, we assume that $Ks$ varies only along the $\hat{x}$ direction and is constant along the $\hat{y}$ axis.
%In the simplest case, $Ks$ may be chosen as follows:
%\begin{equation}
%	Ks(x; t) = \beta_p x - \omega_p t
%\end{equation}
Figure \ref{fig:Zs_TV_1D} shows an example of  spatio-temporally modulated impedance surface in different time slots for $Ks(x; t) = \beta_p x - \omega_p t$ and $Z_{-1} = Z_{+1}$. In the next section, we will see that the impedance synthesized by the holographic method will be similar to that presented in Fig.\ref{fig:Zs_TV_1D}.
The progressive wave-like surface pattern travels with speed of $\nu_p = -\partial_t Ks(x) / (\nabla_x Ks(x).\hat{x}) = \omega_p / \beta_p$ along the $\hat{x}$ axis.
Note that, since the boundary condition of the problem varies with time, we must use the wave equation in the time domain to analyze such structures. Thus, for the half-space above the surface the wave equation can be written as:
\begin{equation}
	(\frac{\partial^2}{\partial x^2} + \frac{\partial ^2}{\partial z^2} - \frac{1}{c^2}\frac{\partial ^2}{\partial t^2}) \psi (x, z; t) = 0
	\label{eq:wave_equation}
\end{equation}
where c is the speed of light in vacuum. 
Note that, for a single frequency system without time modulation, assuming that the wave function depends on 
$e^{j\omega t}$, the wave equation results in the Helmholtz equation. However, in the presence of a time-varying boundary condition and the effects of higher order Floquet harmonics, the Helmholtz equation is not valid.
The periodicity of  boundary condition in both the time and space domains requires that the generalized Floquet-wave theory \cite{cassedy1965} be applied. In this case, the wave function ($\psi(x, z; t)$) can be written in terms of higher order modes \cite{cassedy1965}:
\begin{equation}
	\psi(x, z; t) = \sum_n \psi_n e^{j(\omega + n\omega_p)t} e^{-j(\kappa + n\beta_p)x} e^{-jk_{zn}z}
	\label{eq:psi_Floquet}
\end{equation}
where $\omega$ and $\kappa$ are the fundamental mode frequencies in time and space domains, respectively. The coefficient $\psi_n$ indicates the harmonic amplitude and $jk_{zn}$ is the propagation constant of n-indexed harmonic in the $\hat{z}$ direction. Using (\ref{eq:wave_equation}), $k_{zn}$ can be determined as follows:
\begin{equation}
	k_{zn} = \sqrt{\frac{(\omega + n\omega_p)^2}{c^2} - (\kappa + n \beta_p)^2}
\end{equation}
In equation (\ref{eq:psi_Floquet})  $\omega_p$ and $\beta_p$ can be considered as degrees of freedom, of which the values may be determined by the designer. To estimate $\beta_p$, we can use holographic theory for spatially modulated surfaces.
\section*{Holographic technique for leaky-wave metasurfaces}
An effective approach for the  design and synthesis of leaky-wave modulated metasurface antennas is the holographic technique, which was first proposed by Sievenpiper et al.  \cite{sievenpiper2005, fong2010}. In this technique, for a hologram, the required distribution of surface impedance   to generate an object wave with desired direction  (called $\psi_{obj}$) can be obtained as:
\begin{equation}
	Z_s(x, y) = jX_0[1 + M \times \Re \{\psi_{ref}^*(x, y) \psi_{obj}(x, y)\}]
\label{eq:Zs}
\end{equation}
where $\psi_{ref}$ is the reference wave excited by the electromagnetic source and asterisk ($*$) denotes the complex conjugate operation.  This gradient distribution of surface impedance plays the role of holographic interferogram building up the object wave as the reference wave excites it.
In (\ref{eq:Zs}), $X_0$ and M are average surface reactance and modulation depth, respectively. For leaky-wave holograms the modulated depth must be selected small enough (M < 0.6) in order to achieve the specified directive radiations \cite{oraizi2018}. 
If the radiated wave is supposed to be directed in the $\theta = \theta_0$ zenith angle, the corresponding wave (object wave) can be represented by:
\begin{equation}
	\psi_{obj}(x) = e^{-jk_d \sin \theta_0 x} 
\label{eq:psi_obj}
\end{equation}
where $k_d$ indicates the wave-number at the design frequency $f = f_d$.
For a planar wave-front propagating along the $\hat{x}$ axis, the reference wave  can be defined as \cite{oraizi2018}:
\begin{equation}
	\psi_{ref}(x) = e^{-jk_d\sqrt{1 + (\frac{X_0}{\eta_0})^2}x}
	\label{eq:psi_ref}
\end{equation}
where $\eta_0 = 120 \pi$ is the impedance of free space.
Substituting (\ref{eq:psi_obj}) and (\ref{eq:psi_ref}) in (\ref{eq:Zs}) yields
\begin{equation}
	Z_s(x) = jX_0 [1 + M\times \cos (k_d (\sqrt{1 + (\frac{X_0}{\eta_0})^2} - \sin \theta_0)x )]
	\label{eq:Z_s_2}
\end{equation}
Considering the modulation phase and comparing it with the impedance distribution in the previous section, we can conclude that:
\begin{equation}
	Ks(x) = k_d (\sqrt{1 + (\frac{X_0}{\eta_0})^2} - \sin \theta_0) x
\end{equation}
\begin{equation}
	Z_0 = jX_0, \quad Z_{+1} = Z_{-1} = j\frac{X_0 M}{2}
\end{equation}
\begin{equation}
	\beta_p = k_d (\sqrt{1 + (\frac{X_0}{\eta_0})^2} - \sin \theta_0)
	\label{eq:beta_p}
\end{equation}

Time dependence can be applied to the modulation phase in different ways. However, for simplicity, we assume that the temporal variation of the modulation is linear. Therefore, for dynamic structures, $Ks(x; t)$ may be redefined as follows:
\begin{equation}
	Ks(x; t) = k_d (\sqrt{1 + (\frac{X_0}{\eta_0})^2} - \sin \theta_0) x - \omega_p t
\end{equation}
\section*{Calculation of propagation constant for asymptotic case}
If the surface wave excited by the launcher is in the form of $TM_0$ mode, the x and z components of electric field ($E_x$ and $E_z$) and the y component of magnetic field ($H_y$) may contribute to the propagation along the impedance boundary condition. Using the generalized Floquet-wave expansion we have:
\begin{equation}
	E_x(x, z; t) = \sum_n E_n e^{j(\omega + n\omega_p)t} e^{-j(\kappa + n\beta_p)x} e^{-jk_{zn}z}
	\label{eq:E_x}
\end{equation}
\begin{equation}
	H_y(x, z; t) = \sum_n H_n e^{j(\omega + n\omega_p)t} e^{-j(\kappa + n\beta_p)x} e^{-jk_{zn}z}
	\label{eq:H_y}
\end{equation}
From Maxwell's equations the z component of electric field can be obtained:
\begin{equation}
	E_z(x, z; t) = -\sum_n \frac{(\kappa + n\beta_p)H_n}{(\omega + n\omega_p)\epsilon_0} e^{j(\omega + n\omega_p)t} e^{-j(\kappa + n\beta_p)x} e^{-jk_{zn}z}
	\label{eq:E_z}
\end{equation}
In the above expansions, $\kappa$, $E_n$ and $H_n$ are unknown parameters and must be determined.  The n-indexed Floquet mode frequencies may also be defined as follows:
\begin{equation}
	\beta^{(n)} = \kappa + n\beta_p
	\label{eq:beta_n}
\end{equation}
\begin{equation}
	\omega^{(n)} = \omega + n\omega_p
	\label{eq:omega_n}
\end{equation}
The Eigenmode analysis is an appropriate method to calculate the unknowns ($\kappa$, $E_n$ and $H_n$). Substituting (\ref{eq:E_x}) and (\ref{eq:H_y}) in the impedance boundary condition, and comparing the terms with identical modulation phases, leads to
\begin{equation}
	E_n = -jX_0 [H_n + \frac{M}{2}(H_{n+1} + H_{n-1})]
	\label{eq:E_n}
\end{equation} 
Furthermore, the Maxwell equation imposes the following restriction:
\begin{equation}
	\frac{E_n}{H_n} = \frac{k_{zn}}{(\omega + n\omega_p)\epsilon_0}
	\label{eq:E_n_H_n}
\end{equation}
Combining (\ref{eq:E_n}) and (\ref{eq:E_n_H_n}) yields:
\begin{equation}
	H_{n+1} + D_n H_n + H_{n-1} = 0 \quad n = 0, \pm 1, \pm 2, ...
	\label{eq:dispersion_tv}
\end{equation}
where:
\begin{equation}
	D_n = \frac{2}{M}[1 - \frac{jk_{zn}}{X_0(\omega + n \omega_p)\epsilon_0}]
	\label{eq:D_n}
\end{equation}
Equation (\ref{eq:dispersion_tv}) relates the magnitude of n-indexed harmonic to (n+1) and (n-1)-indexed modes, which form a set of infinite equations with infinite unknowns. 
These equations resemble the dispersion relations in the pure-space modulated case \cite{oliner1959}, except that $\omega ^{(n)} = \omega + n\omega_p$ is used instead of $\omega$. 
In order to solve this system of equations, we must truncate the number of modes for analysis.

\begin{figure}[ht]
\centering
\begin{subfigure}[]{0.32\textwidth}
	\includegraphics[width=\linewidth]{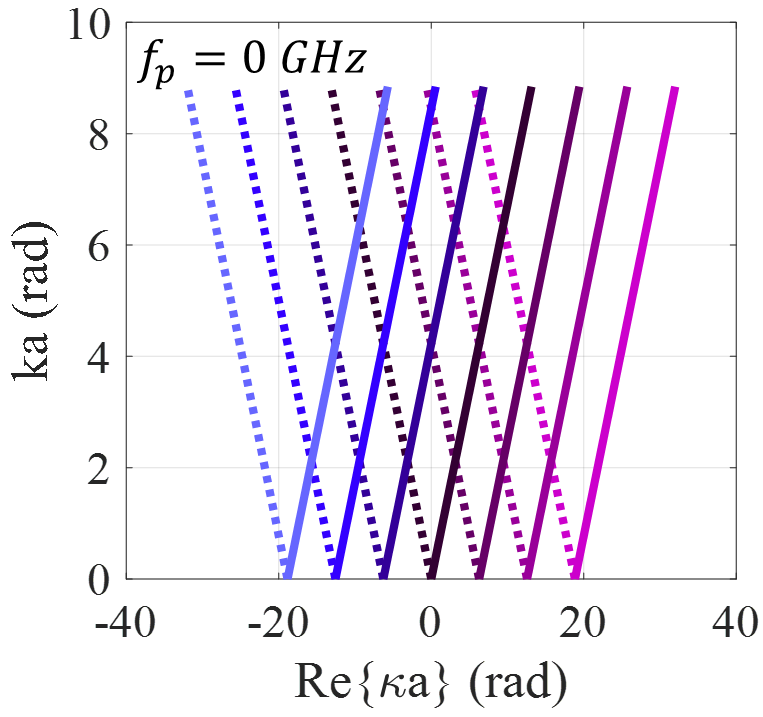}
	\caption{}
\end{subfigure}
\begin{subfigure}[]{0.32\textwidth}
	\includegraphics[width=\linewidth]{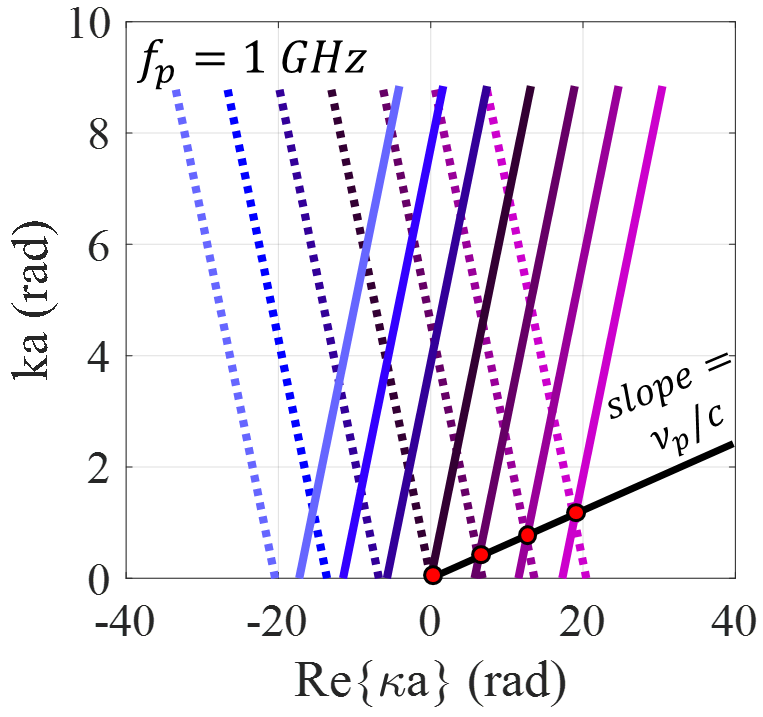}
	\caption{}
\end{subfigure}
\begin{subfigure}[]{0.32\textwidth}
	\includegraphics[width=\linewidth]{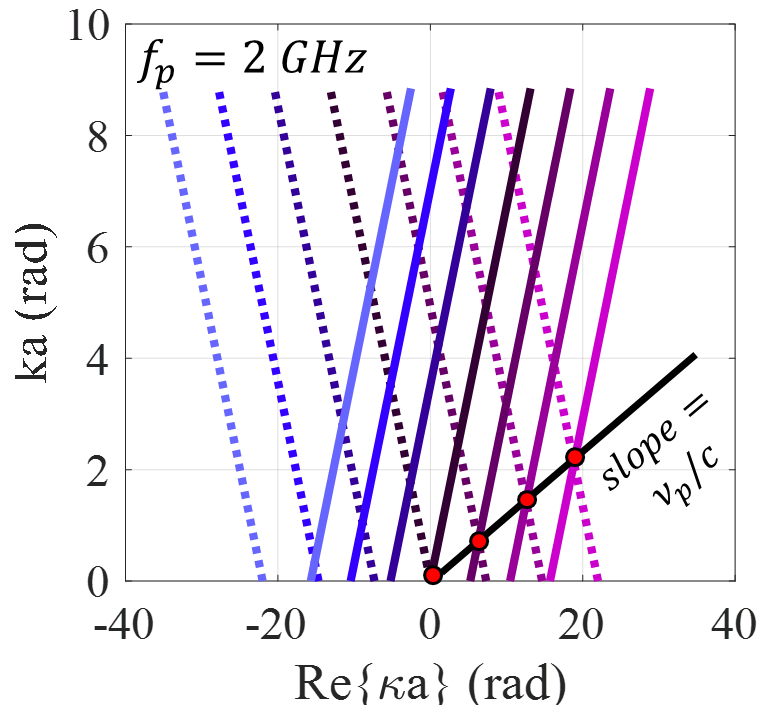}
	\caption{}
\end{subfigure}
\caption{Dispersion curves for different pumping frequencies. (a) $f_p = $, (b) $f_p = 1 GHz$, and (c) $f_p = 2 GHz$.}
\label{fig:dispersion_um}
\end{figure}

For the asymptotic case, if the modulation index tends to zero ($M\rightarrow 0$), the numerator of fraction in equation (\ref{eq:D_n}) needs to be zero to achieve nontrivial solutions. In this case we have:
\begin{equation}
	1 - \frac{j\sqrt{\frac{(\omega + n\omega_p)^2}{c^2} - (\kappa + n\beta_p)^2}}{X_0(\omega + n \omega_p) \epsilon_0} = 0
\end{equation}
Therefore, the set of asymptotic curves for $\kappa$ can be obtained from the following equation:
\begin{equation}
	\kappa = -nk_d (\sqrt{1 + (\frac{X_0}{\eta_0})^2} - \sin \theta_0) \pm \frac{\omega + n\omega_p}{c} \sqrt{1 + (\frac{X_0}{\eta_0})^2}  \quad n = 0, \pm 1, \pm 2, ...
\end{equation}  
Figure \ref{fig:dispersion_um} shows the asymptotic dispersion curves for the cases of $f_p = 0$, $f_p = 1 GHz$ and $f_p = 2 GHz$. 
The solid and dashed curves represent forward and backward solutions, respectively.  These curves can be used as initial guesses for the calculation of eigenstates of the modulated case. Observe that, injection of temporal  variations into the modulated impedance function imposes some asymmetry on the dispersion curves, leading to a nonreciprocal response. 
As shown in Fig. \ref{fig:dispersion_um}, in the presence of temporal modulation, the Brillouin diagram is parallel to the line with slope of $\nu_p / c$ and the period of Brillouin zone (the distance between adjacent red points in Fig. \ref{fig:dispersion_um}) is $\sqrt{\beta_p^2 + (\omega_p/c)^2}$. Observe that increasing the temporal pumping frequency results in the increase of the degree of asymmetry.
This property  will be described in more detail in the following section.

\section*{Determination of propagation constant in the presence of space-time modulation}
\begin{figure}
	\centering
	\begin{subfigure}{0.5\textwidth}
		\includegraphics[width = \linewidth]{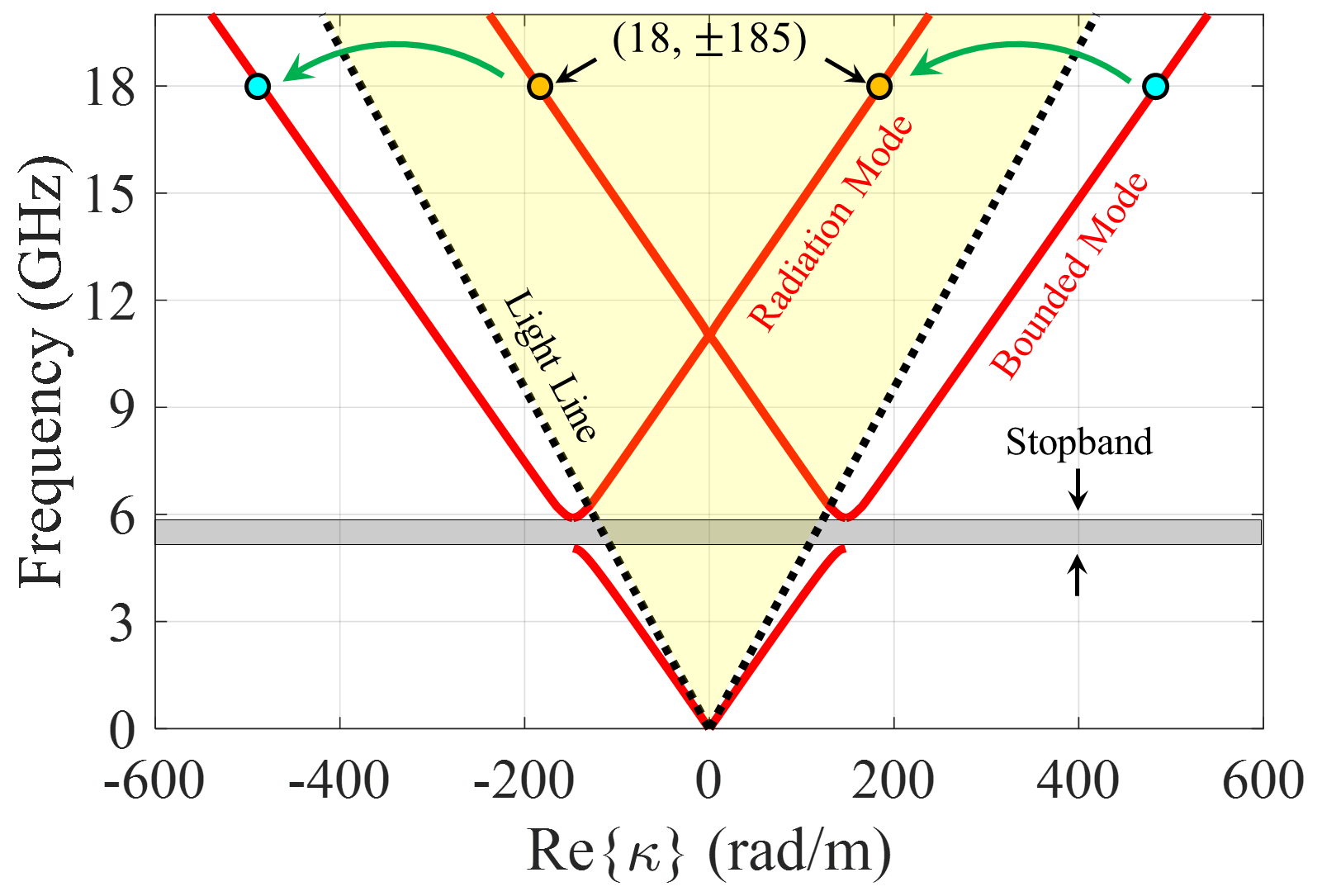}
		\caption{}
	\end{subfigure}
	\begin{subfigure}{0.4\textwidth}
		\includegraphics[width = \linewidth]{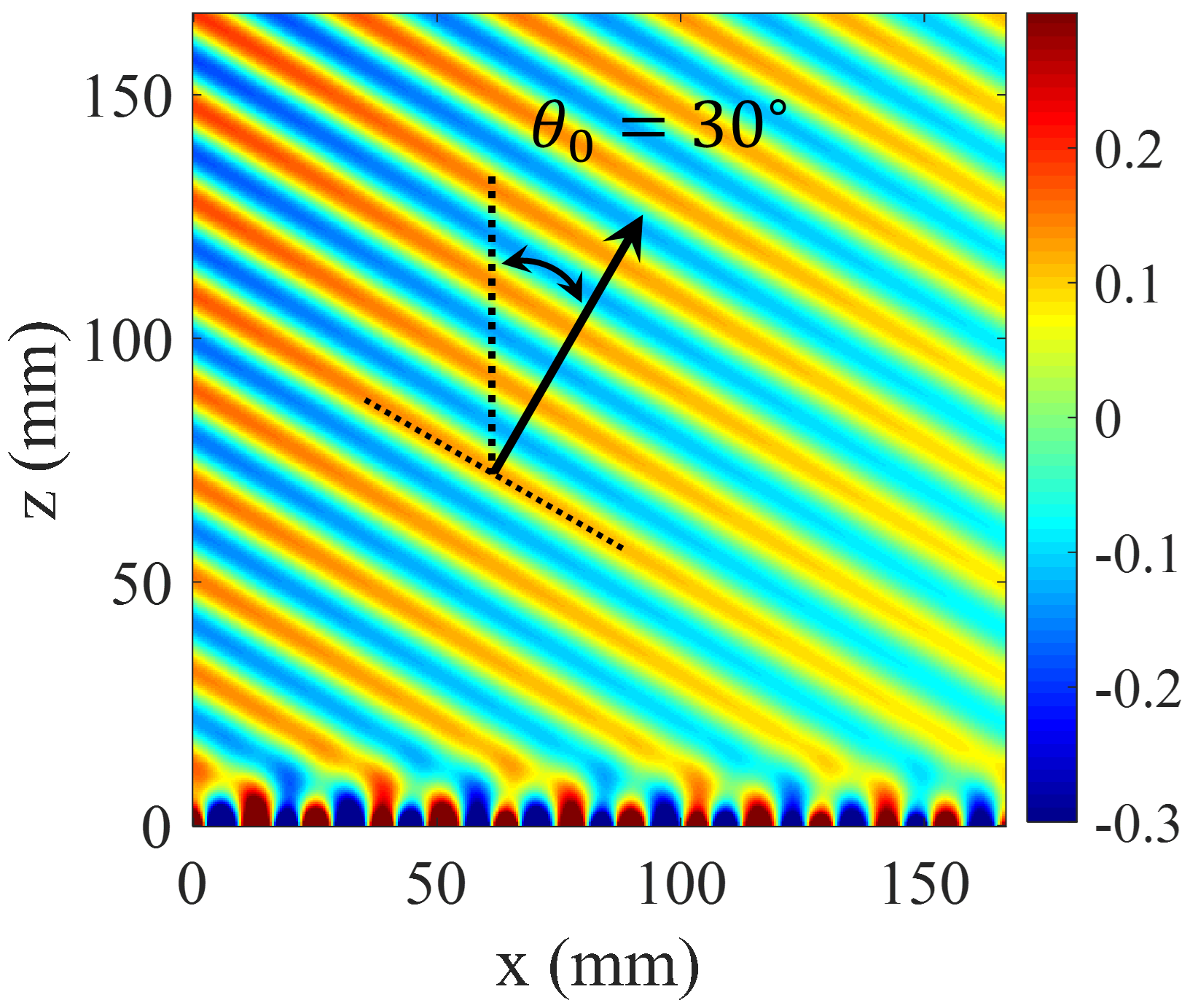}
		\caption{}
	\end{subfigure}
	\caption{(a) Dispersion diagram of proposed hologram in the pure-space modulation mode ($f_p = 0\, GHz$). (b) corresponding bounded and radiation fields at 18 GHz.}
	\label{fig:dispersion_static}
\end{figure}

\begin{figure}
	\centering
	\begin{subfigure}{0.46\textwidth}
		\includegraphics[width = \textwidth]{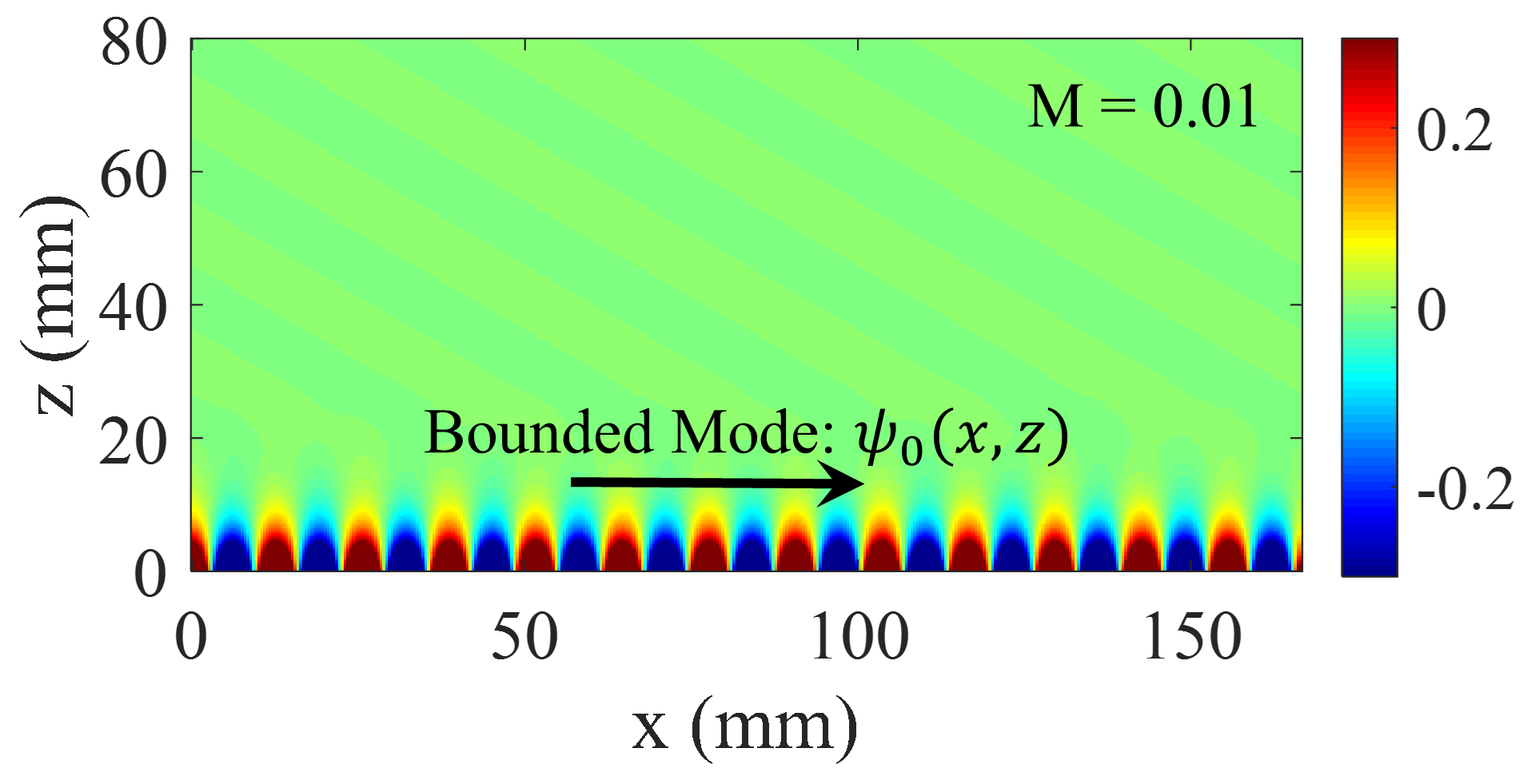}
		\caption{}
	\end{subfigure}
	\begin{subfigure}{0.46\textwidth}
		\includegraphics[width = \textwidth]{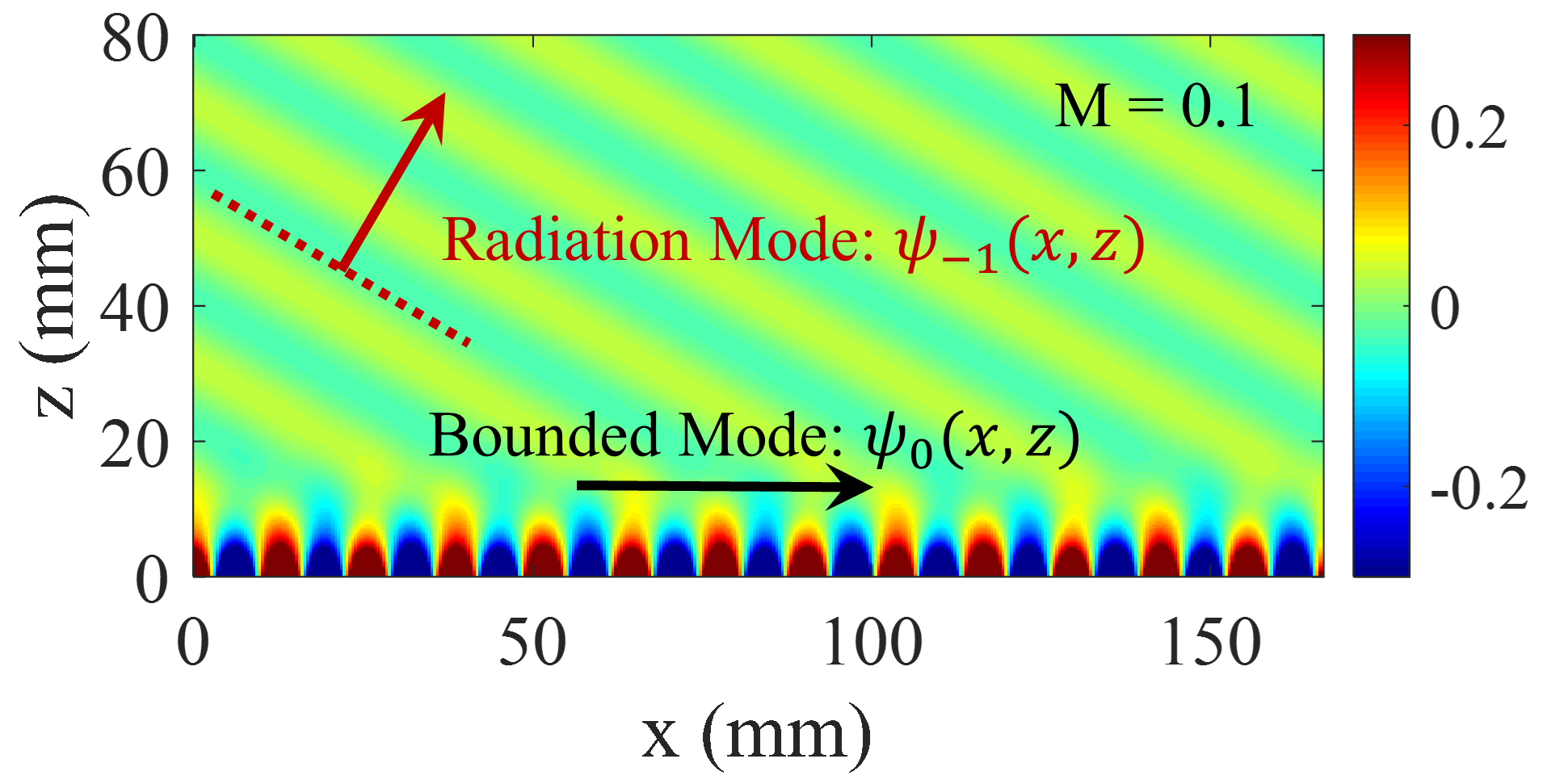}
		\caption{}
	\end{subfigure}
	\begin{subfigure}{0.46\textwidth}
		\includegraphics[width = \textwidth]{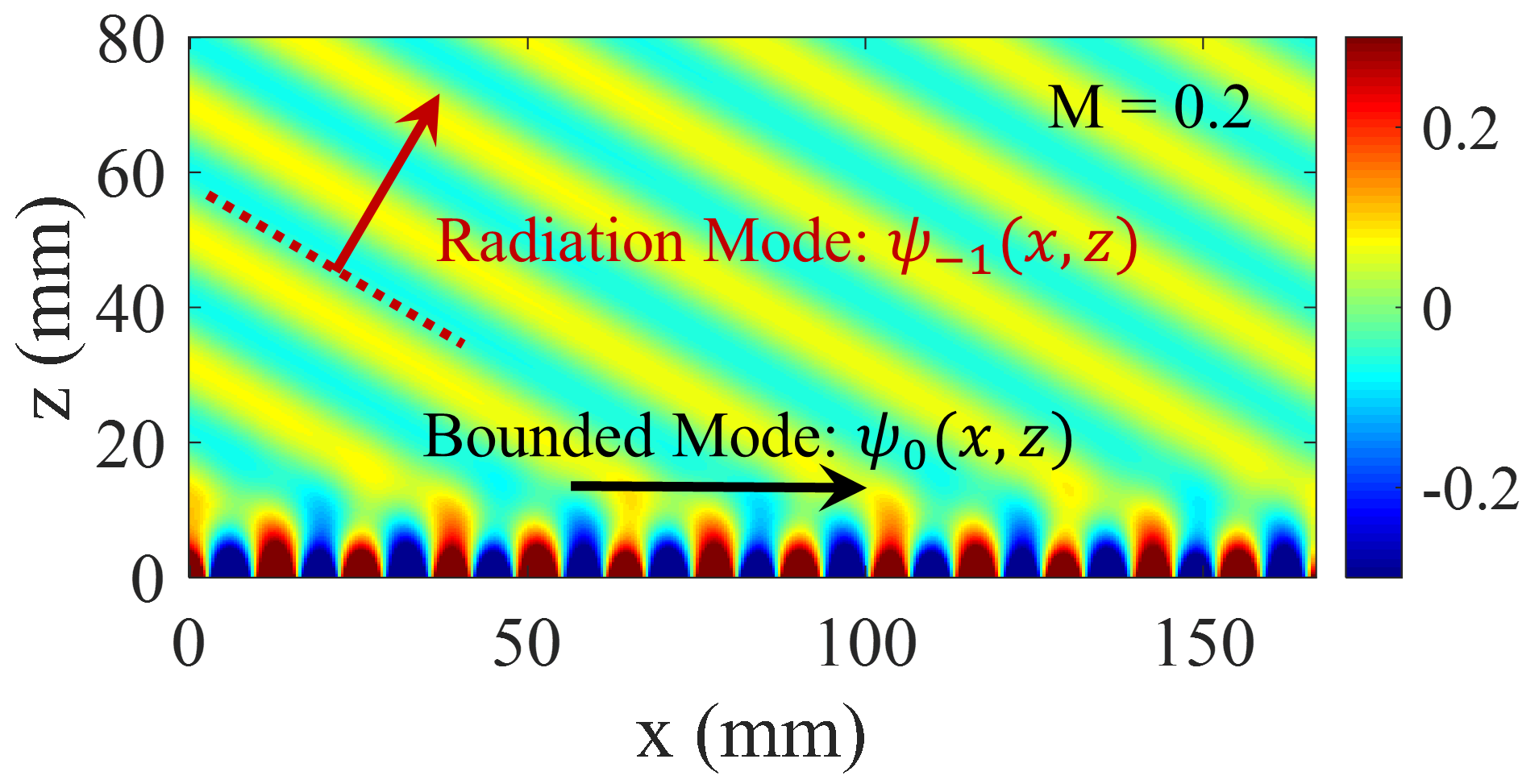}
		\caption{}
	\end{subfigure}
	\begin{subfigure}{0.46\textwidth}
		\includegraphics[width = \textwidth]{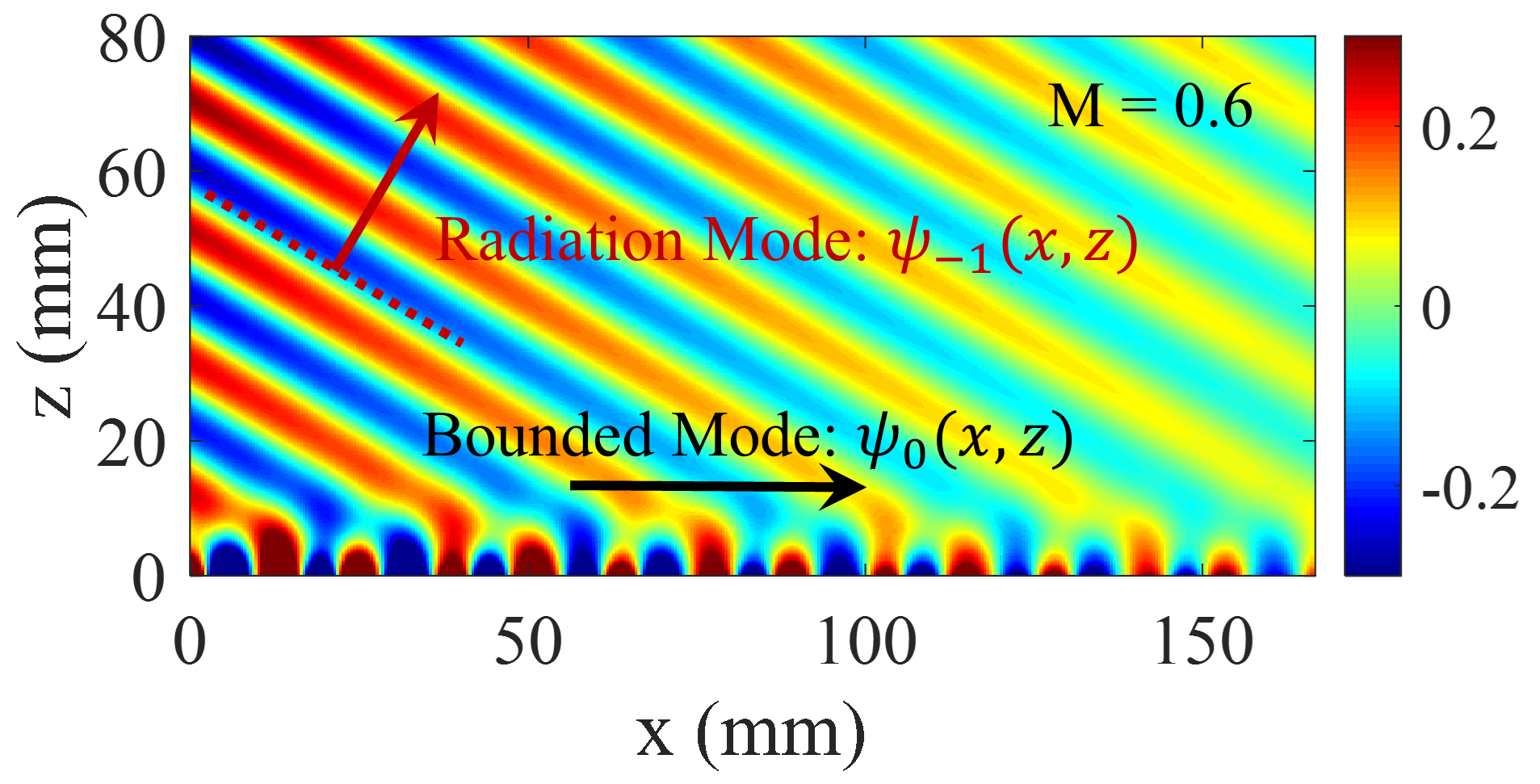}
		\caption{}
	\end{subfigure}
	\caption{Field distribution for different values of modulation depth. (a) $M = 0.01$, (b) $M = 0.1$, (c) $M = 0.2$, and (d) $M = 0.6$.}
	\label{fig:M_0}
\end{figure}

As discussed in the previous section, 
The infinite system of equations in (\ref{eq:dispersion_tv}) should be solved for the determination of the exact values of eigenstates in the presence of space-time modulation, which is achieved by safely truncating the number of modes without generating of a significant error. 
 In reference \cite{casaletti2019} it has been shown that considering only three Floquet modes (namely $n = 0, \pm 1$) results in sufficiently accurate solutions. Therefore, the following equations will be formed:
\begin{equation}
	H_0 + D_{-1} H_{-1} = 0
	\label{eq:D_minus}
\end{equation}
\begin{equation}
	H_1 + D_0 H_0 + H_{-1} = 0
	\label{eq:D_0}
\end{equation}
\begin{equation}
	D_1 H_1 + H_0 = 0
	\label{eq:D_plus}
\end{equation}
Combining (\ref{eq:D_minus})-(\ref{eq:D_plus}), yields:
\begin{equation}
	(D_0 - \frac{1}{D_{-1}} - \frac{1}{D_1}) H_0 = 0
\end{equation}
To obtain nontrivial solutions, it is required to have:
\begin{equation}
	D_0 - \frac{1}{D_{-1}} - \frac{1}{D_1} = 0
\end{equation}
Therefore, we can rewrite the dispersion equation  as:
\begin{equation}
	\frac{4}{M^2}[1 - \frac{j\sqrt{(\omega/c)^2 - \kappa^2}}{X_0 \omega \epsilon_0}] = \frac{1}{1 - \frac{j\sqrt{(\omega - \omega_p)^2/c^2 - (\kappa - \beta_p)^2}}{X_0(\omega - \omega_p) \epsilon_0}}   +    \frac{1}{1 - \frac{j\sqrt{(\omega + \omega_p)^2/c^2 - (\kappa + \beta_p)^2}}{X_0(\omega + \omega_p) \epsilon_0}} 
\end{equation}
Its solution determines the complex values of  wave-number at a given frequency. Figure \ref{fig:dispersion_static}a shows the dispersion diagram of hologram for the case of $f_p = 0\, GHz$. The design frequency is  18 GHz, and the beam angle at 18 GHz is set at $\theta_0 = 30^\circ$. The average impedance ($X_0$)  and modulation index (M) are selected as $0.8\eta_0$ and 0.4. Therefore, the corresponding impedance may be obtained as:
\begin{equation}
	Z_s(x) = j 0.8 \eta_0 [1  + 0.4 \cos (377.25(\sqrt{1 + 0.8^2} - 0.5)x)] = j301.59[1 + 0.4\cos (294.49 x)]
\end{equation}
The total magnetic field at 18 GHz is also plotted in Fig. \ref{fig:dispersion_static}b, which show that, in the transmission mode, the coupling occurs from the surface to radiation mode with the radiation angle of 30 degrees. Also, in the reception mode, the maximum coupling occurs at the incoming angle of 30 degrees, which confirms the accuracy of the design.
In holographic antennas, the coupling strength depends on the modulation depth (M). Figure \ref{fig:M_0} shows the total fields for different values of modulation depths. Observe that for lower values of $M$, most of the power is bound to the surface. The power coupling is increased by enhancing the modulation depth.

\begin{figure}
	\centering
	\begin{subfigure}{0.45\textwidth}
		\includegraphics[width = \textwidth]{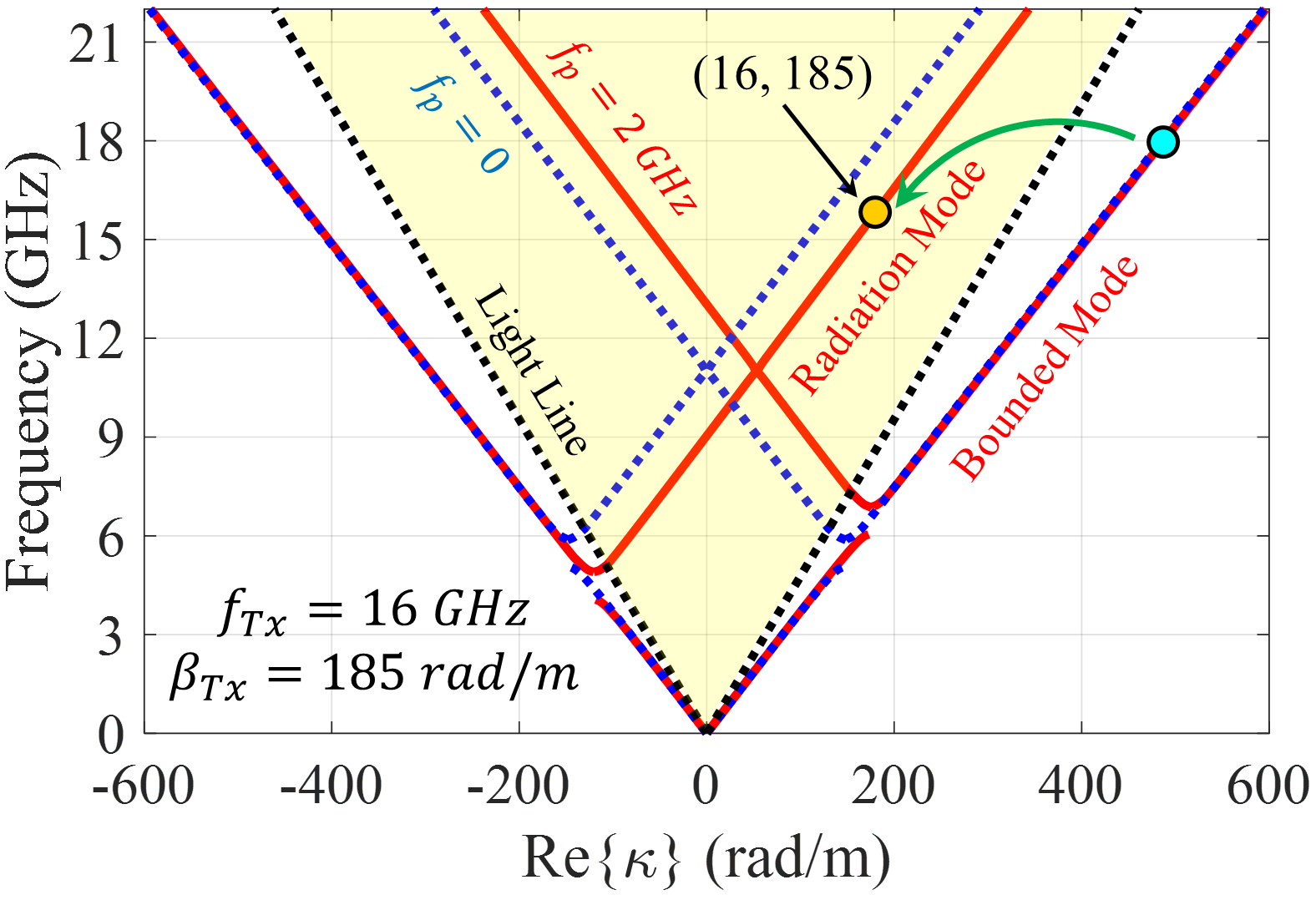}
		\caption{}
	\end{subfigure}
	\begin{subfigure}{0.45\textwidth}
		\includegraphics[width = \textwidth]{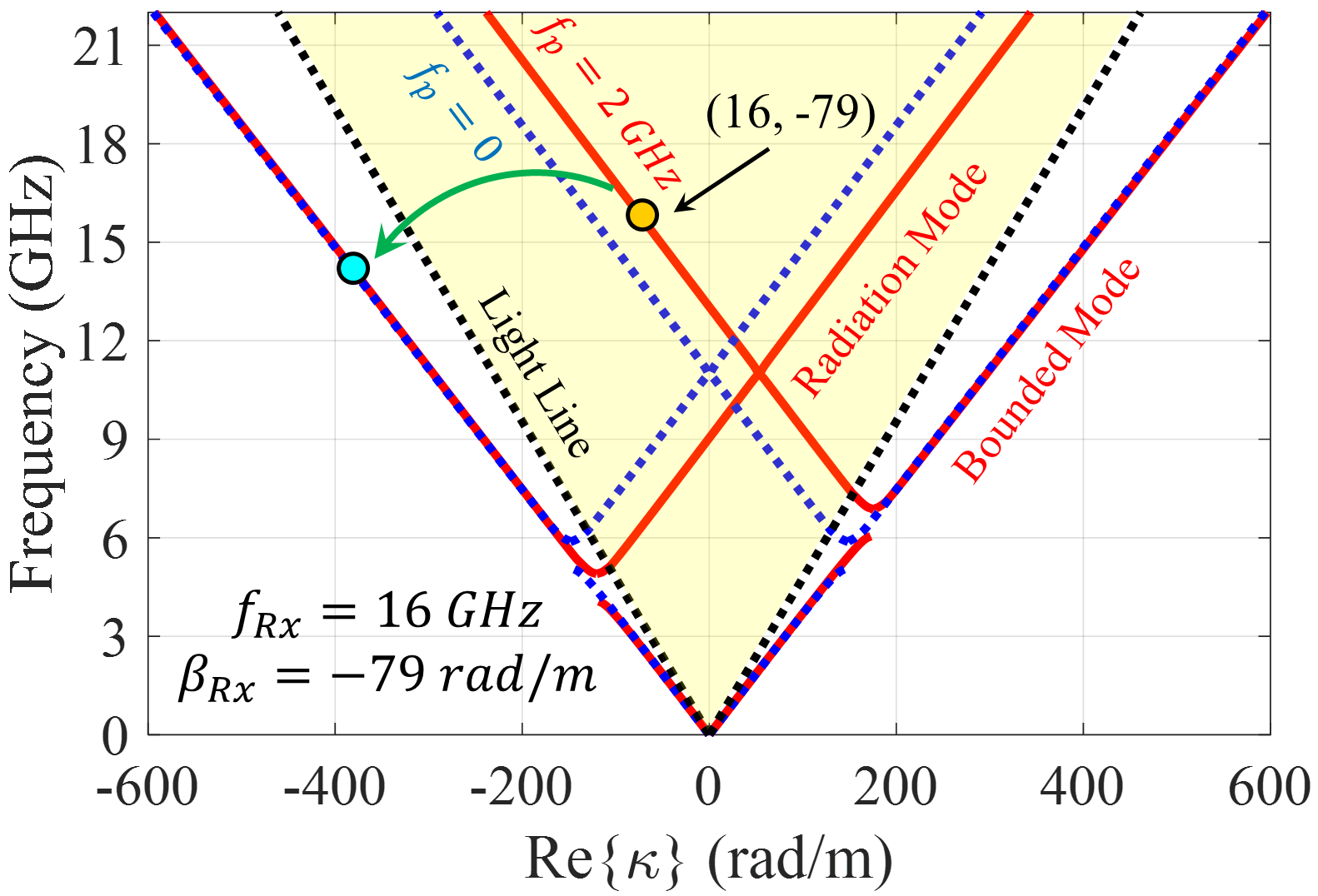}
		\caption{}
	\end{subfigure}
	\caption{Dispersion diagram of the  proposed spatiotemporally modulated hologram for the cas of $f_p = 2 \, GHz$ in (a) transmission mode, and (b) receiving mode.}
	\label{fig:dispersion_Tx_Rx}
\end{figure}

\begin{figure}
	\centering
	\begin{subfigure}{0.45\textwidth}
		\includegraphics[width = \textwidth]{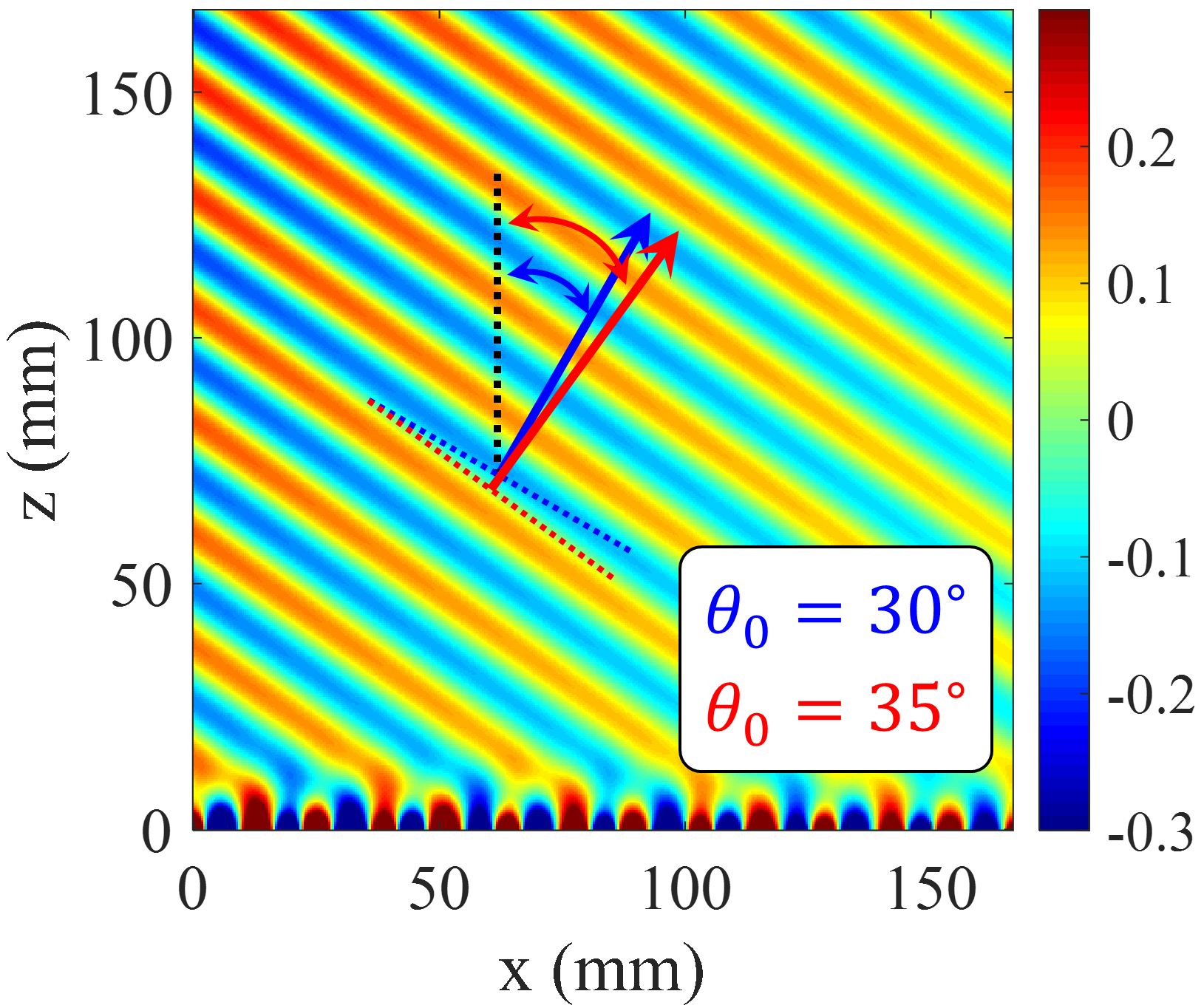}
		\caption{}
	\end{subfigure}
	\begin{subfigure}{0.45\textwidth}
		\includegraphics[width = \textwidth]{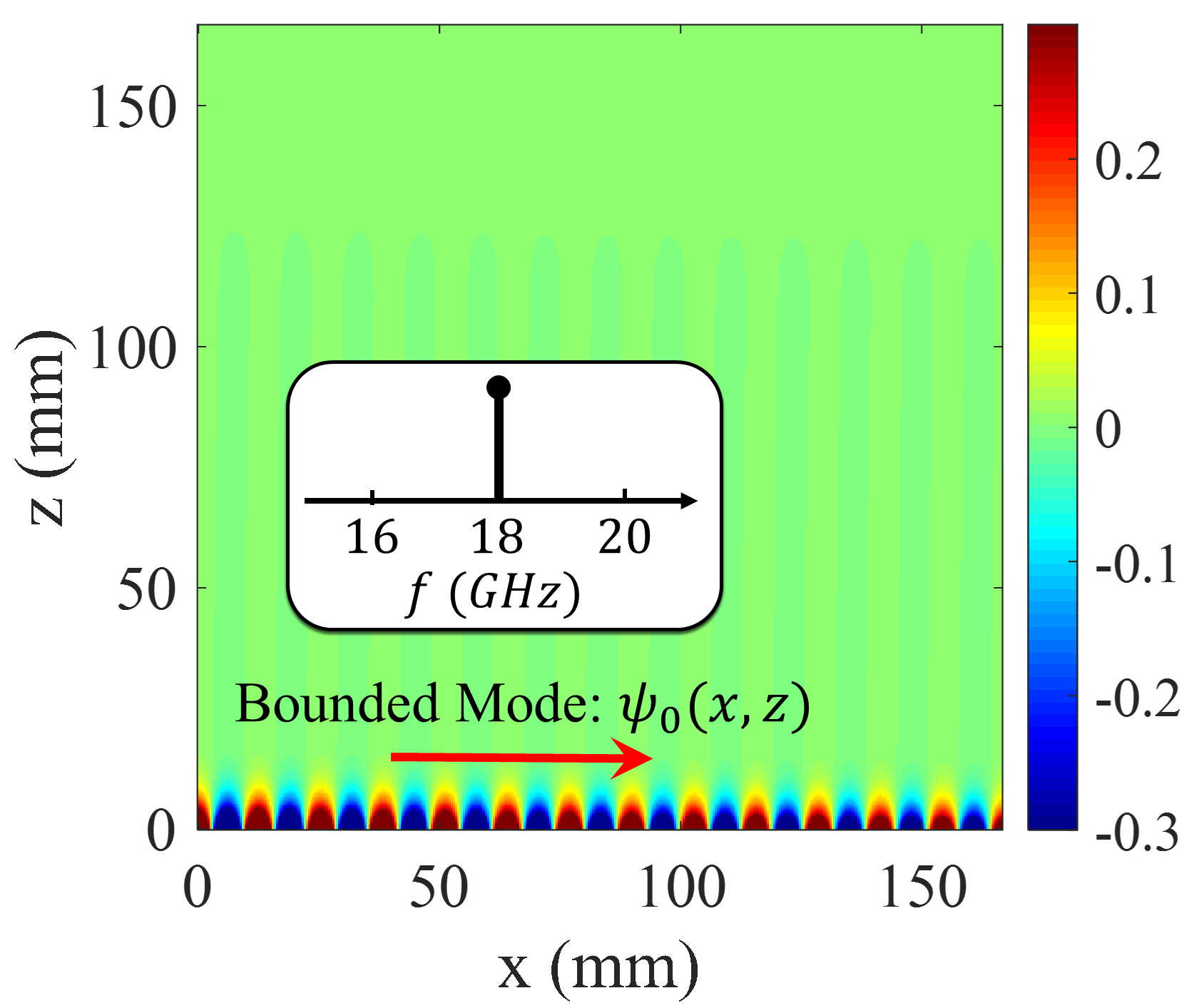}
		\caption{}
	\end{subfigure}
	\begin{subfigure}{0.45\textwidth}
		\includegraphics[width = \textwidth]{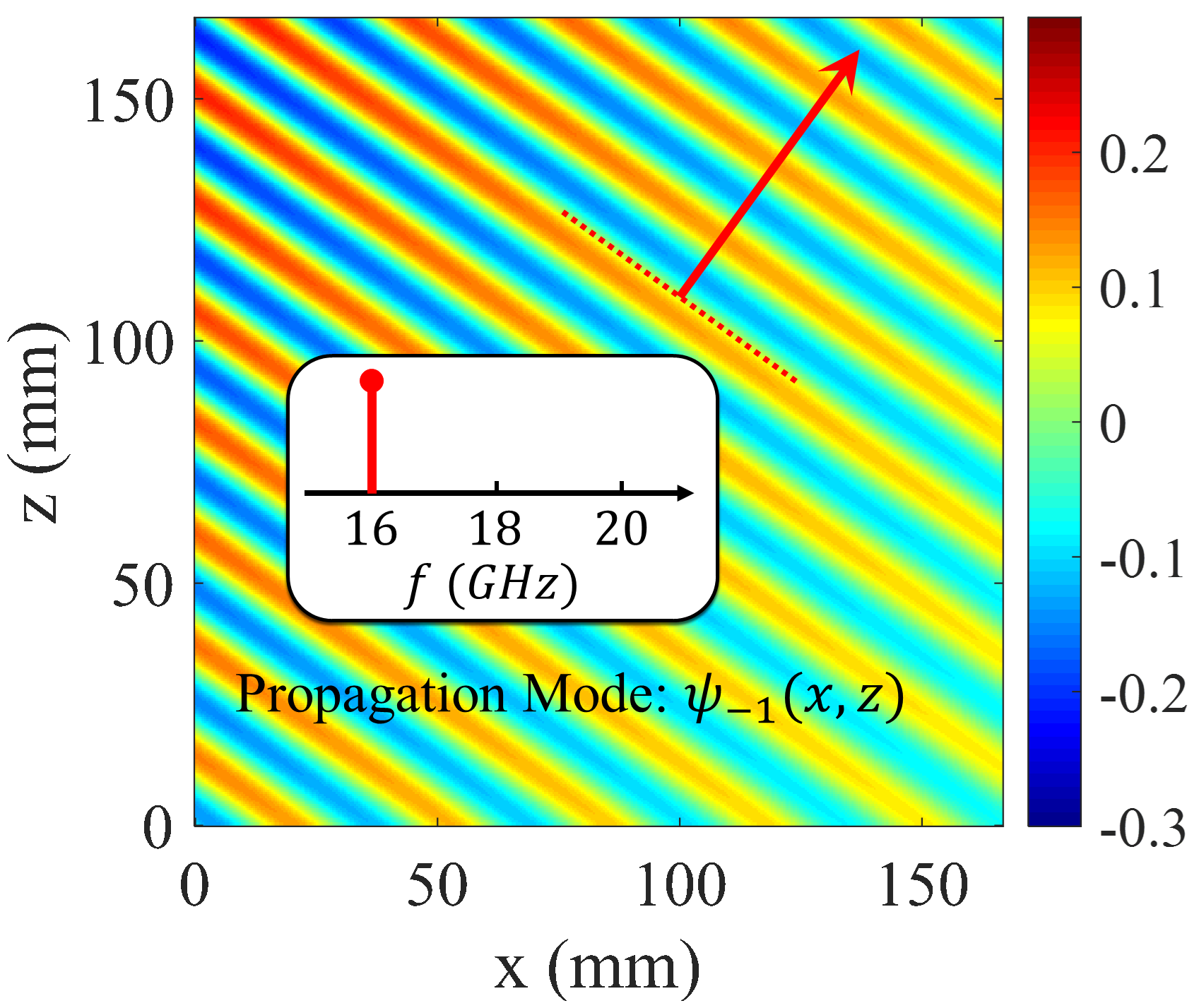}
		\caption{}
	\end{subfigure}
	\begin{subfigure}{0.45\textwidth}
		\includegraphics[width = \textwidth]{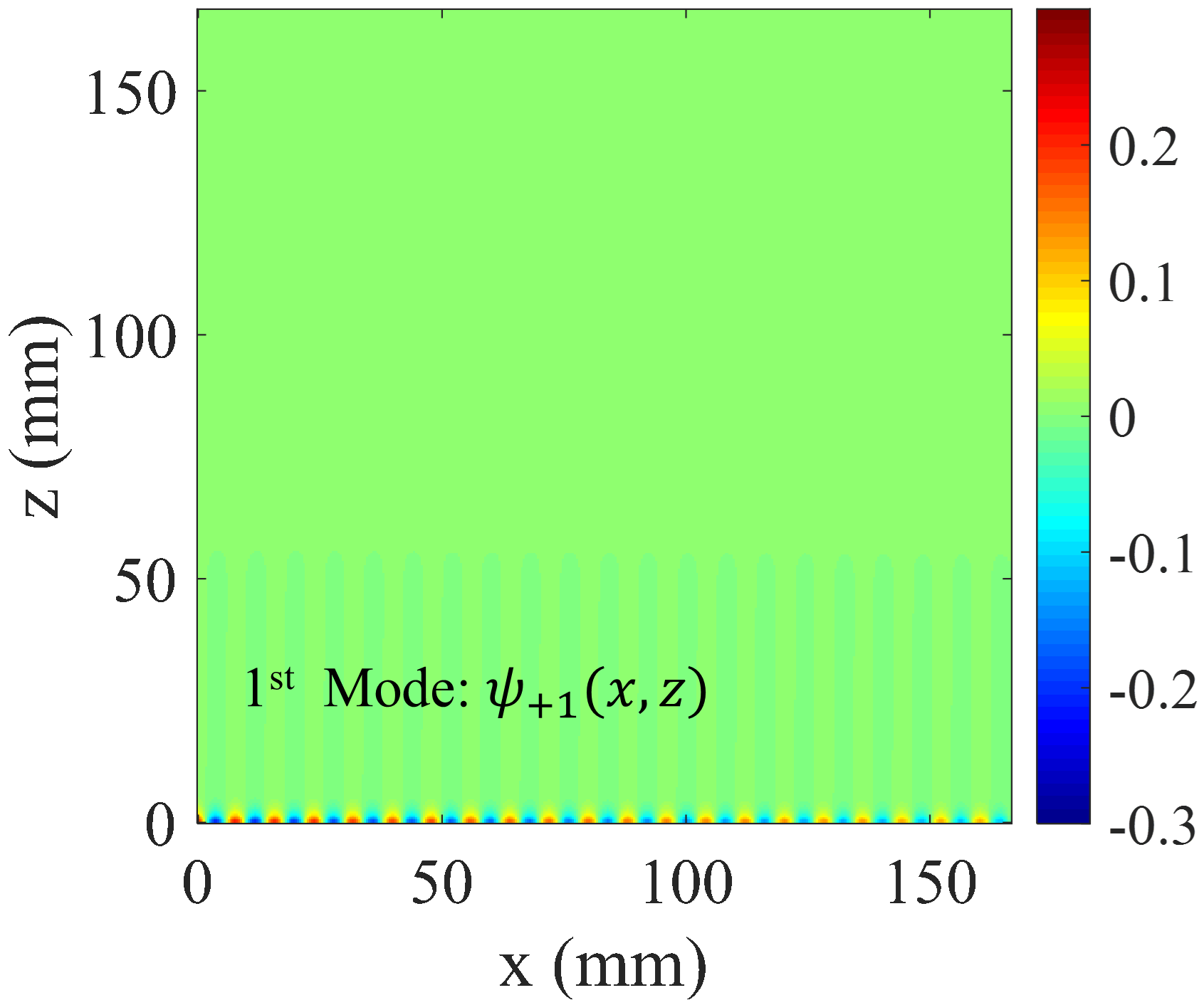}
		\caption{}
	\end{subfigure}
	\caption{Analytical results of the proposed spationtemporally modulated hologram in transmission mode. (a) total field, (b) fundamental mode, (c) radiation mode, and (d) (+1)-index mode. }
	\label{fig:H_Tx}
\end{figure}

\begin{figure}
	\centering
	\begin{subfigure}{0.45\textwidth}
		\includegraphics[width = \textwidth]{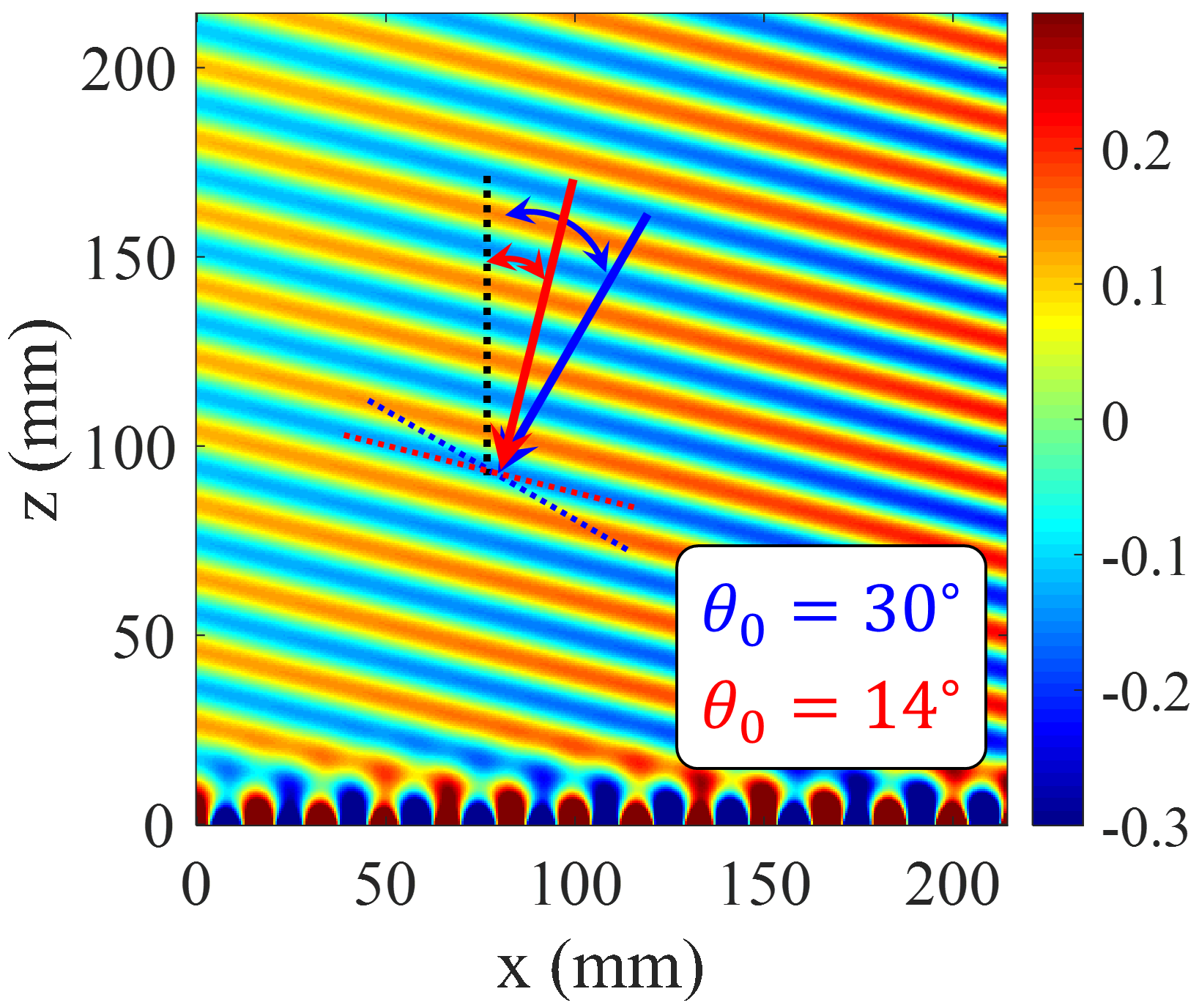}
		\caption{}
	\end{subfigure}
	\begin{subfigure}{0.45\textwidth}
		\includegraphics[width = \textwidth]{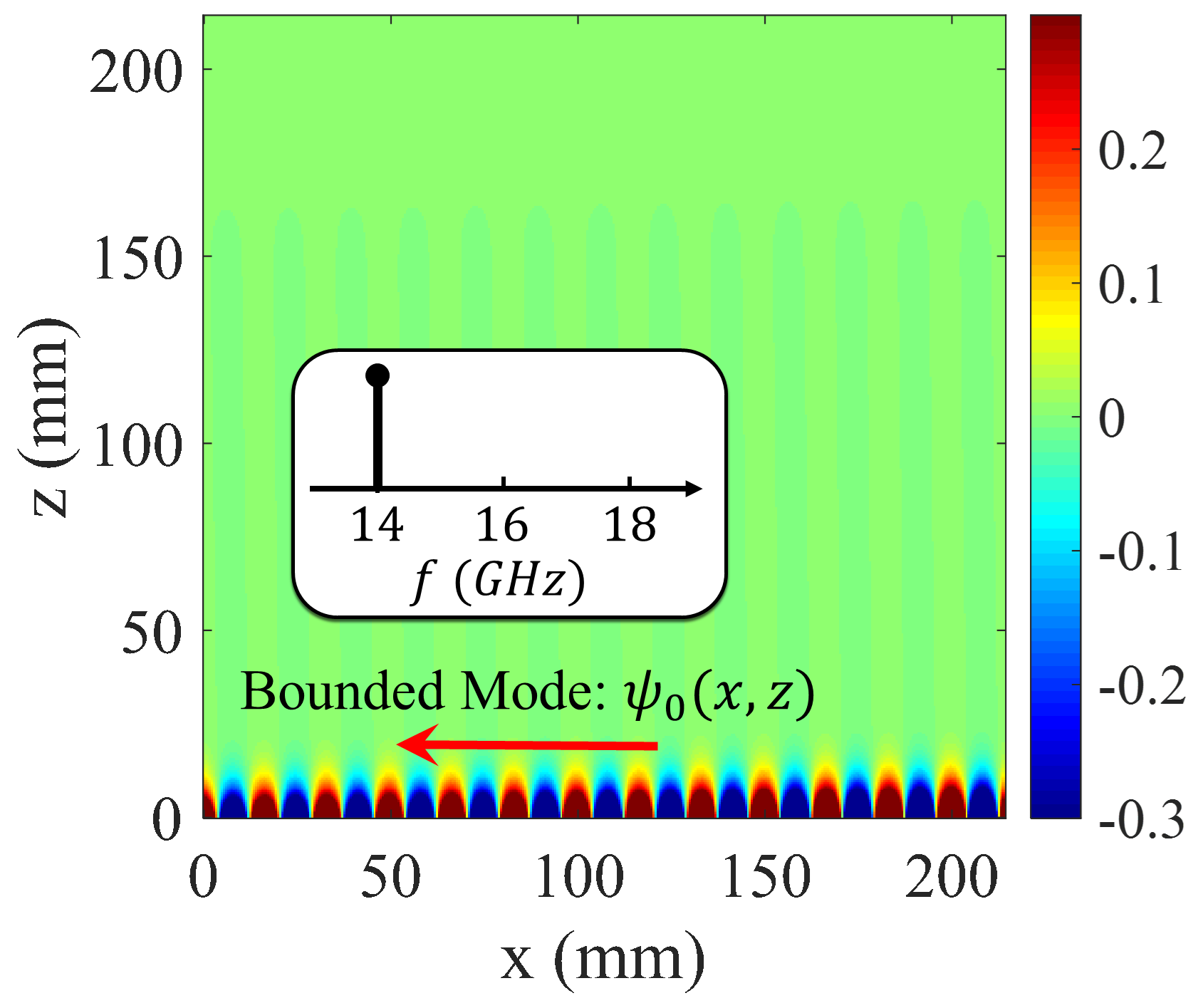}
		\caption{}
	\end{subfigure}
	\begin{subfigure}{0.45\textwidth}
		\includegraphics[width = \textwidth]{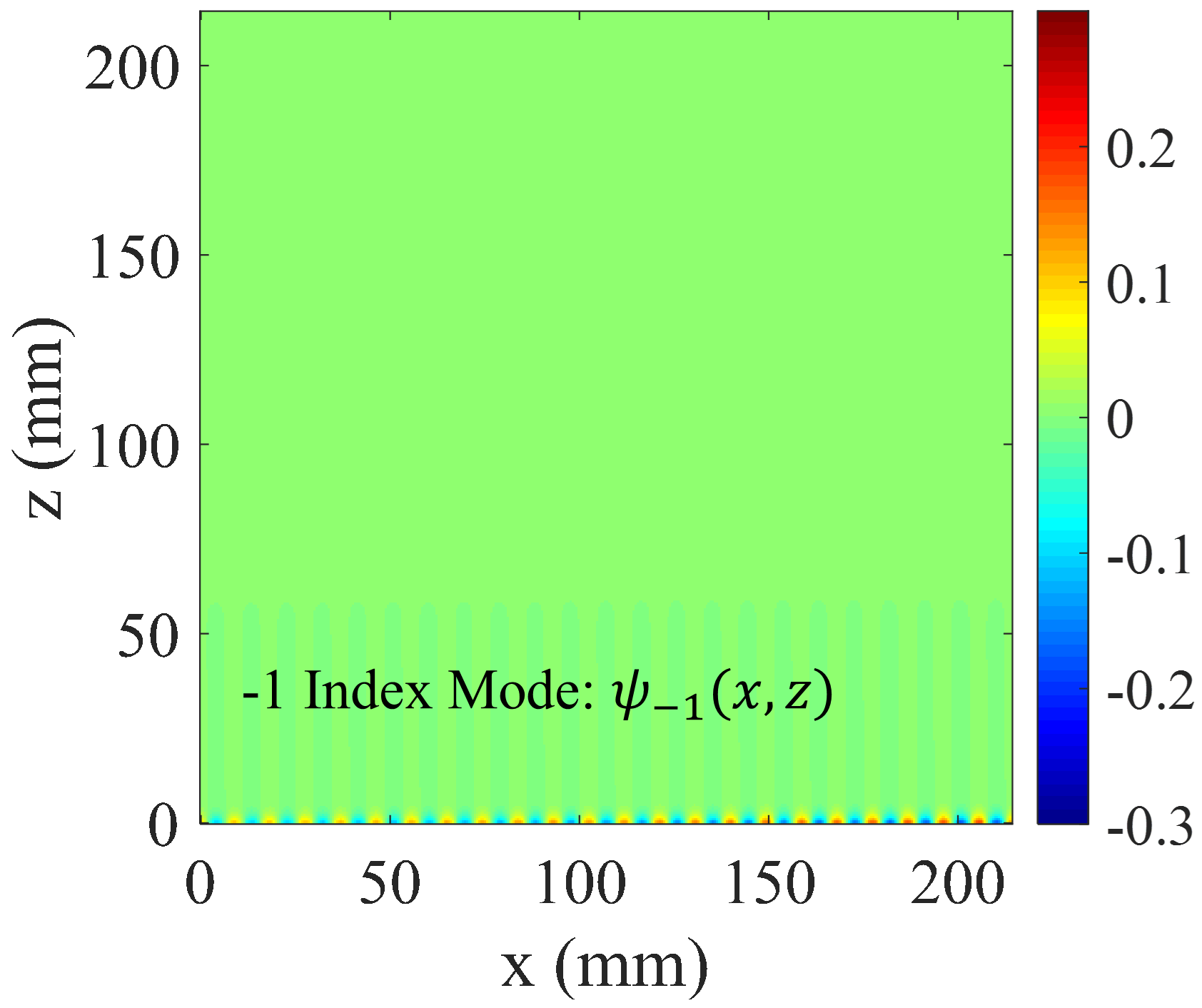}
		\caption{}
	\end{subfigure}
	\begin{subfigure}{0.45\textwidth}
		\includegraphics[width = \textwidth]{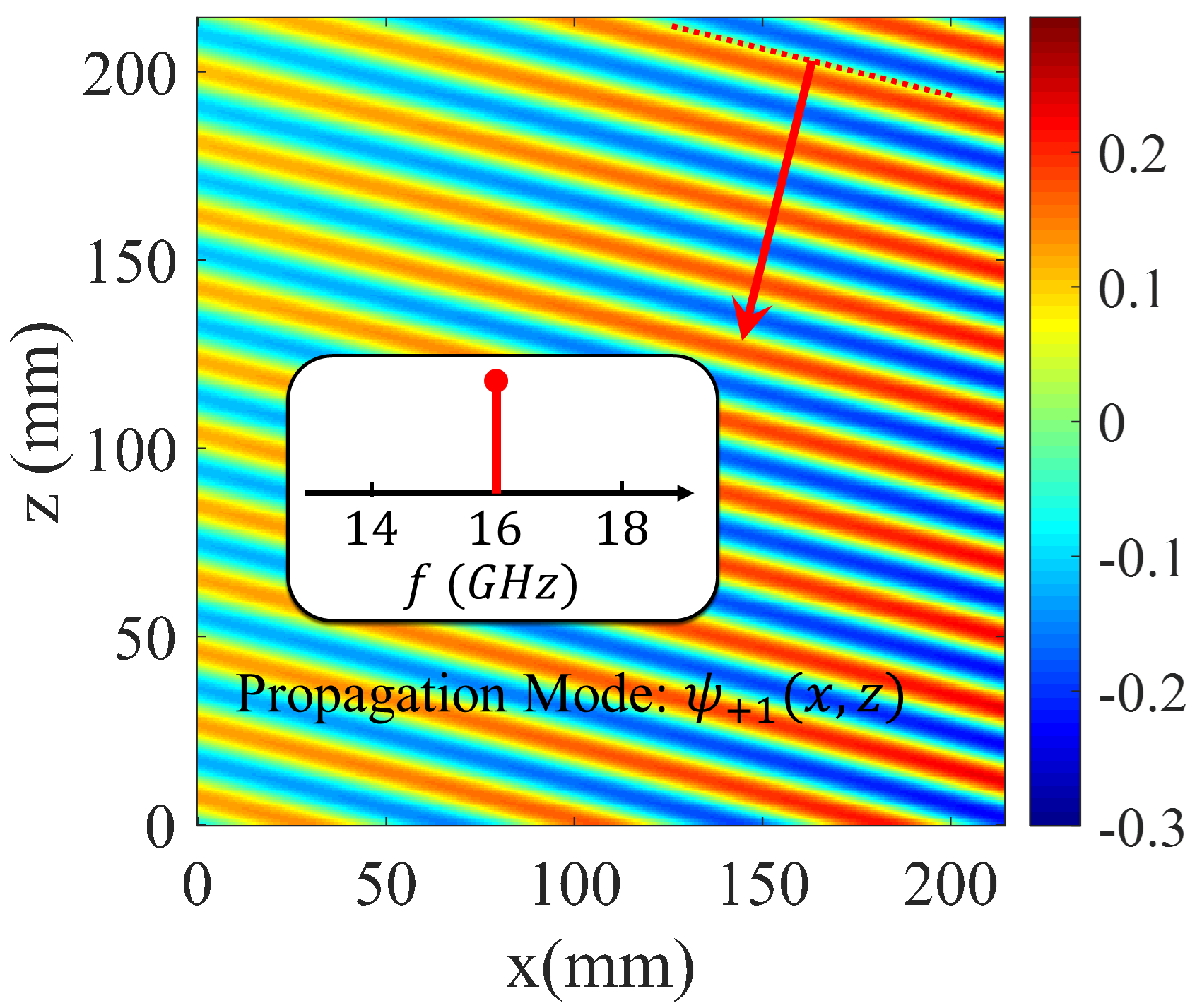}
		\caption{}
	\end{subfigure}
	\caption{Analytical results of the proposed spationtemporally modulated hologram in receiving mode. (a) total field, (b) fundamental mode, (c) (-1)-index mode, and (d) radiation mode. }
	\label{fig:H_Rx}
\end{figure}

Injecting temporal modulation into the designed hologram, imposes some modifications to the dispersion diagram.
Figure \ref{fig:dispersion_Tx_Rx} shows the dispersion curves for the case of $f_p = 2 \, GHz$. Its impedance distribution has the following form:
\begin{equation}
	Z_s(x; t) = j301.59[1 + 0.4 \cos (294.49x - 4\pi \times 10^9 t)]
\end{equation}
Observe that, in the presence of temporal modulation, the curves corresponding to negative and positive harmonics  move down and up respectively. This displacement is proportional to the pumping frequency ($f_p$). As the pumping frequency increases, the dispersion curve for $n = -1$ tends towards the light-line, indicating that the radiation mode tends towards the end-fire direction. Also for $n = +1$, the radiation curve gets closer to the vertical axis.
To investigate nonreciprocity, we first consider the antenna operation  in transmission mode. Suppose that the excitation frequency of the antenna is 18 GHz, which operates at the fundamental mode (surface wave). According to the dispersion diagram in Fig. \ref{fig:dispersion_Tx_Rx}a, its corresponding phase constant is equal to $\beta^{(0)} = 488 \, (rad/m)$. Furthermore, using (\ref{eq:beta_n}) the phase constant for radiation mode ((-1)-indexed mode) can be obtained as $\beta^{(-1) } = \beta^{(0)} - \beta_p \approx 185 \, (rad/m)$. Observe in the dispersion diagram  that the corresponding frequency at this phase constant is 16 GHz (namely $f^{(-1)} = f^{(0)} - f_p$), which shows the Doppler-shift effect in the proposed antenna. According to the calculated phase constant, the beam orientation is directed at 35 degrees. In Fig. \ref{fig:H_Tx} the analytical results of total magnetic field and Floquet-modes of $n = 0, \pm 1$ are plotted separately. 
Figure \ref{fig:H_Tx} indicates that a strong coupling can be observed between 0-indexed and (-1)-indexed Floquet modes. In this case, most of the power is divided into the fundamental  and (-1)-indexed modes, but the (+1)-indexed mode has a small portion of the power.
The radiation mode is oriented in the direction of 35 degrees. The blue and red arrows in Fig. \ref{fig:H_Tx}a indicate the beam directions for pure-space and space-time modulated cases,  respectively. 
When the antenna is in the reception mode, for the 16 GHz  incoming wave, the corresponding phase constant is entirely different from that of the transmission case. 
Using (\ref{eq:omega_n}), for (+1)-indexed mode with frequency of 16 GHz, the 0-indexed mode propagates at $f^{(0)} = f^{(+1)} - f_p = 14 GHz$.
In this case, the phase constant of (+1)-indexed mode is equal to $\beta^{(+1)} \approx -79 \, (rad/m)$ and the angle at which the maximum power is coupled to the surface is 14 degrees (as shown in Fig. \ref{fig:H_Rx}). Note that the surface is excited at 14 GHz instead of 16 GHz. 

To summarize our analysis we can conclude that, for the antenna  in the transmission mode, the maximum coupling between the surface and radiation modes occurs at the angle of 35 degrees, whereas for the reception case the radiation and surface modes have the maximum coupling at the incoming angle of 14 degrees . Therefore, the antenna does not behave the same in the transmission and reception modes, which confirms its nonreciprocal behavior. 
\section*{Beam scanning properties}
\begin{figure}
	\centering
	\includegraphics[width = 0.55\textwidth]{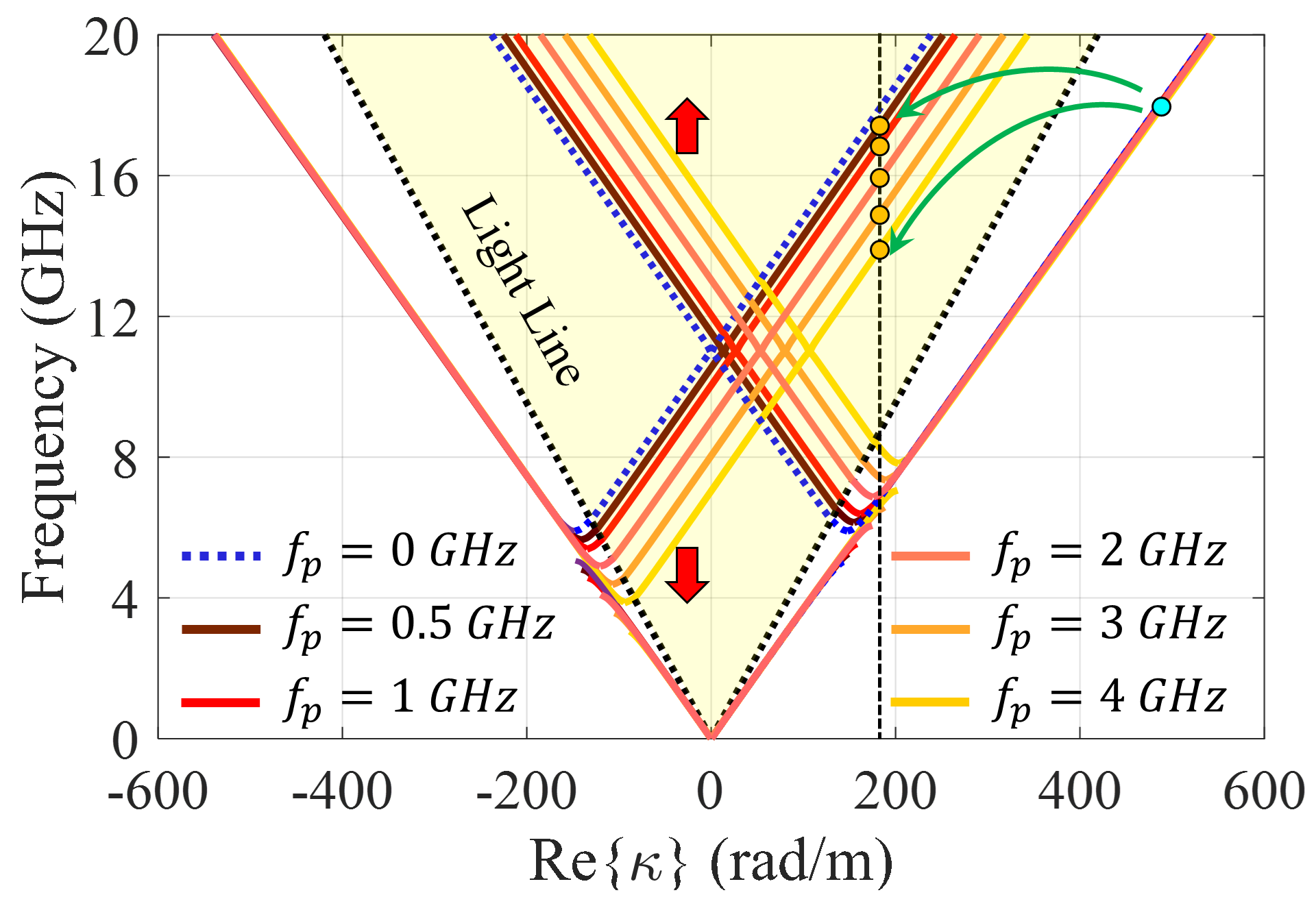}
	\caption{Dispersion diagram of proposed hologram for different values of $f_p$.}
	\label{fig:dispersion_fp}
\end{figure}

\begin{figure}
	\centering
	\begin{subfigure}{0.46\textwidth}
		\includegraphics[width = \textwidth]{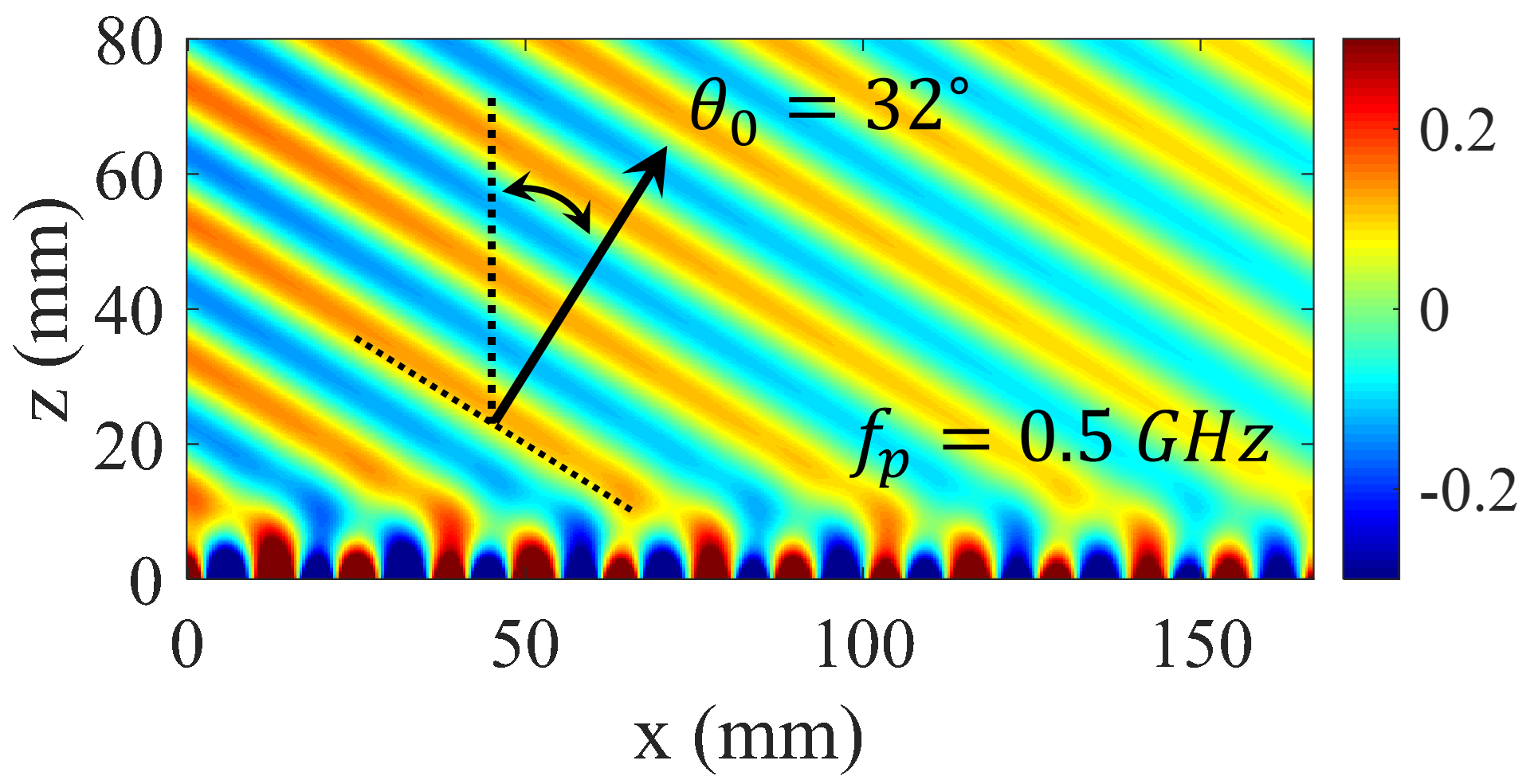}
		\caption{}
	\end{subfigure}
	\begin{subfigure}{0.46\textwidth}
		\includegraphics[width = \textwidth]{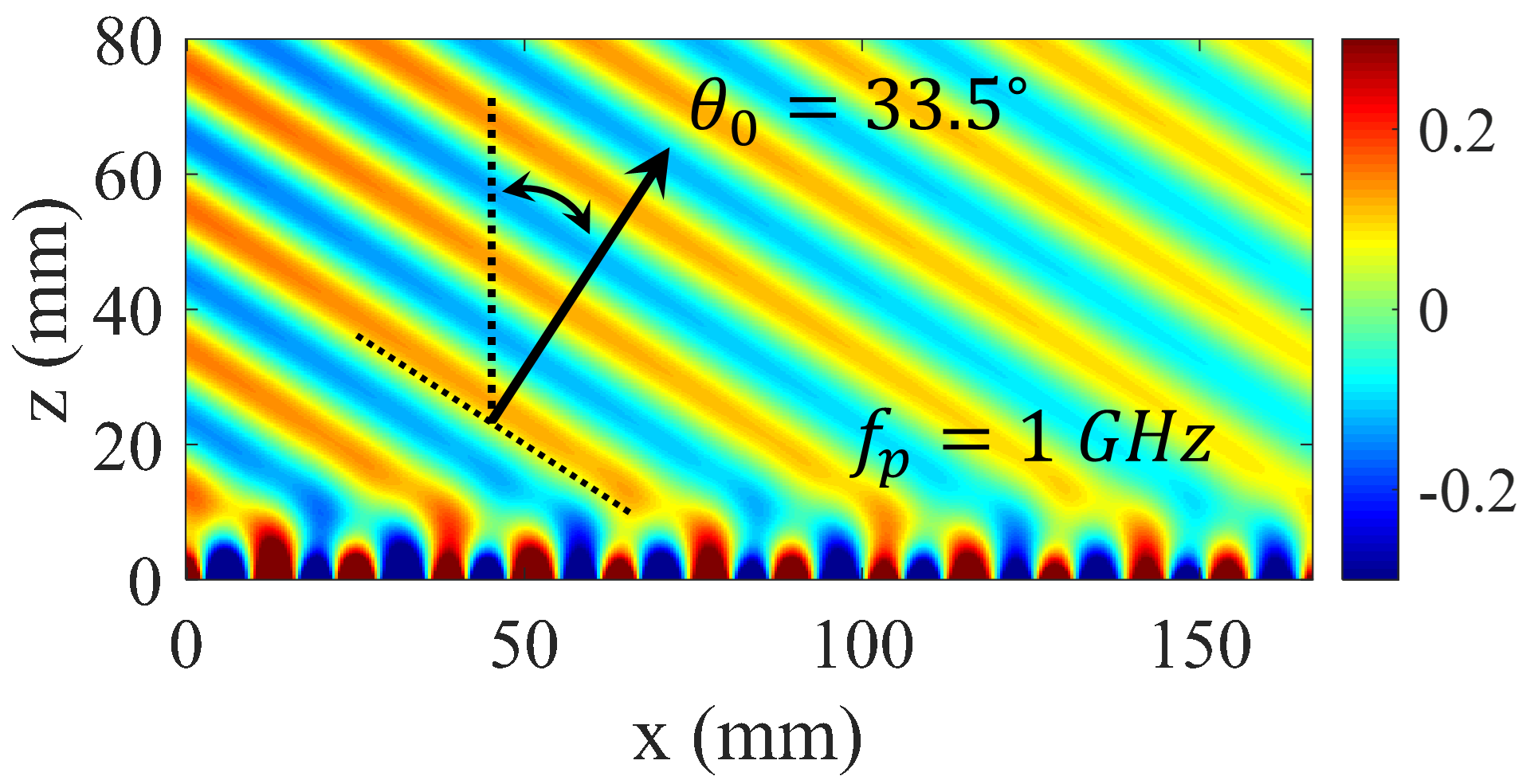}
		\caption{}
	\end{subfigure}
	\begin{subfigure}{0.46\textwidth}
		\includegraphics[width = \textwidth]{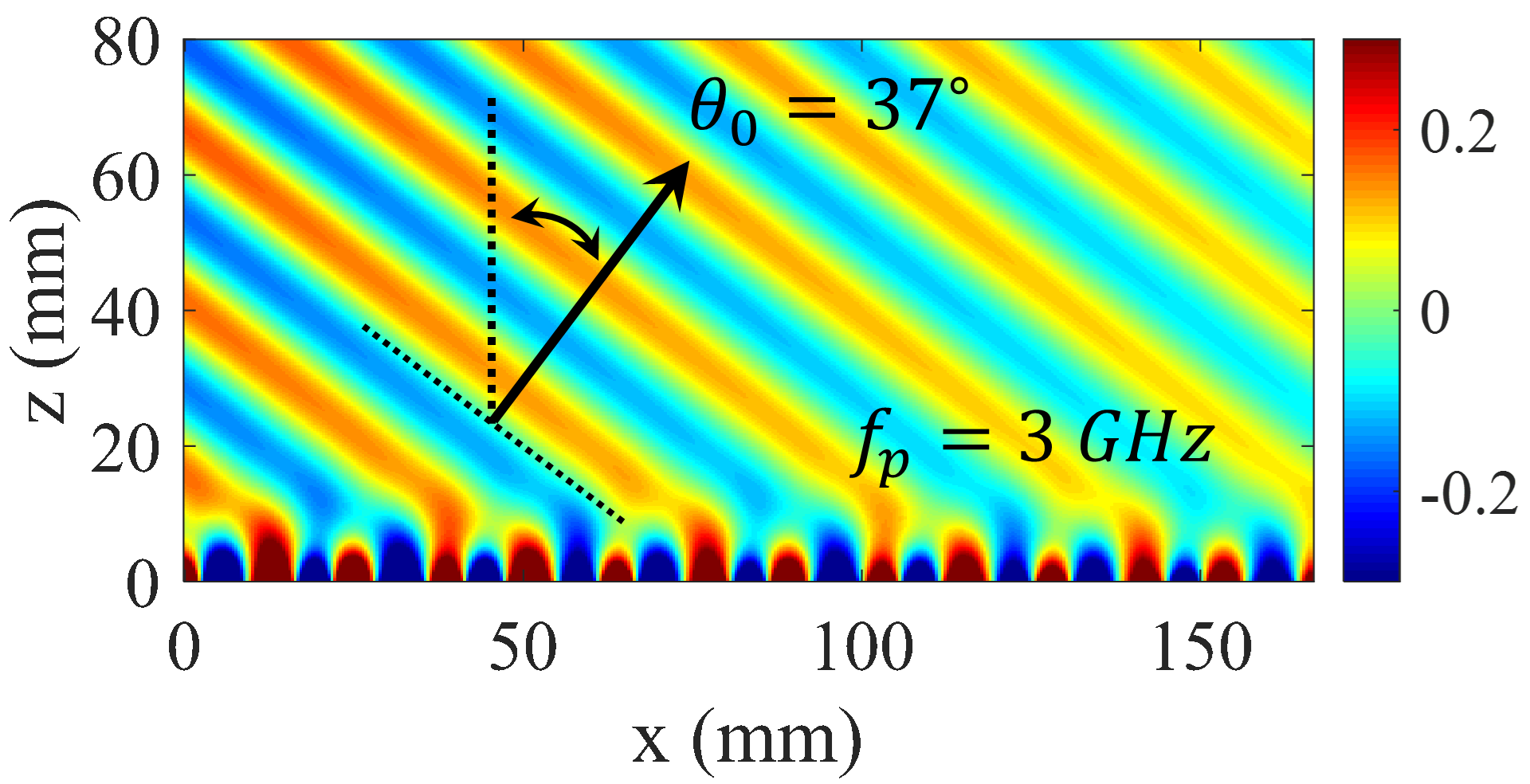}
		\caption{}
	\end{subfigure}
	\begin{subfigure}{0.46\textwidth}
		\includegraphics[width = \textwidth]{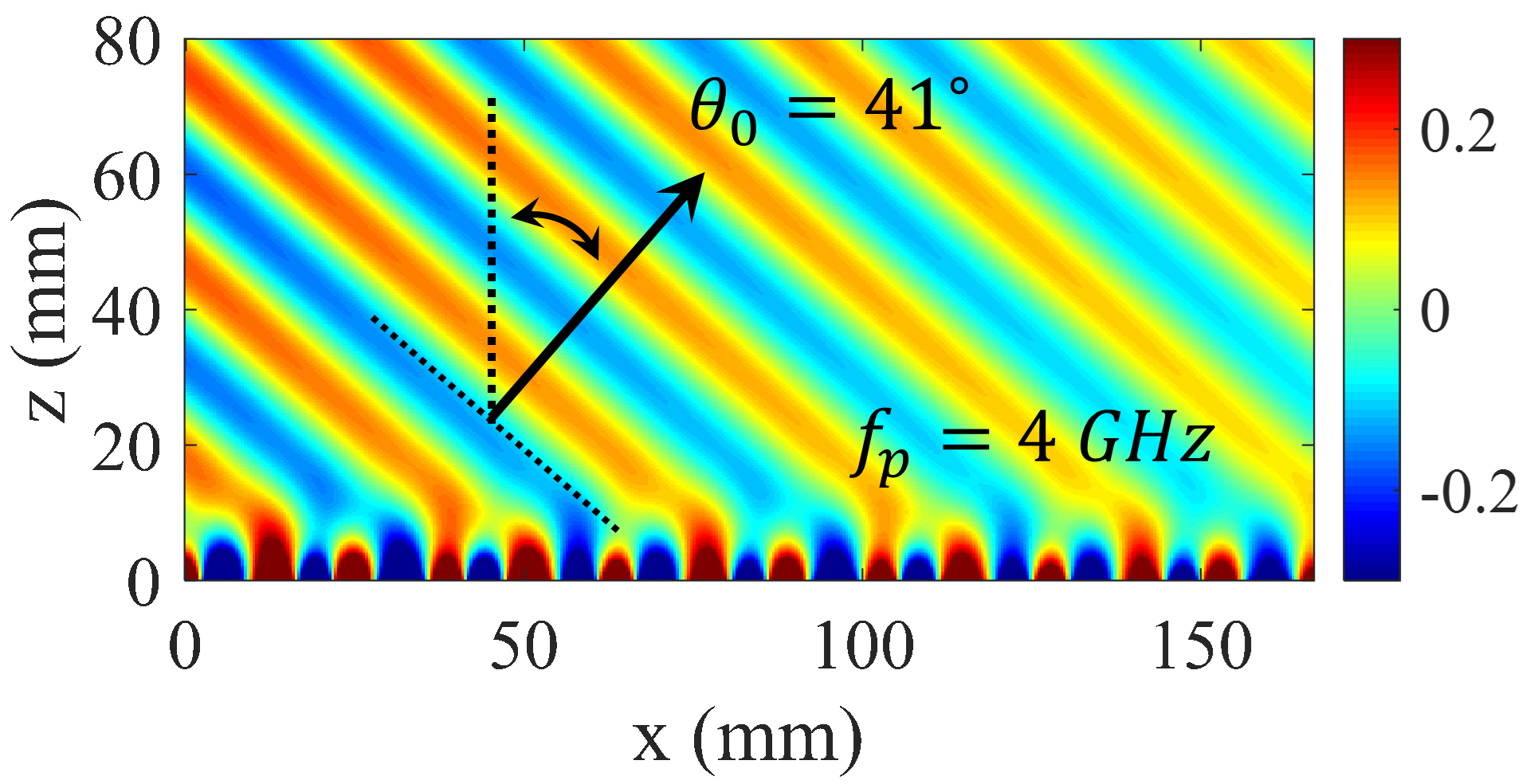}
		\caption{}
	\end{subfigure}
	\caption{Total fields for different values of pumping frequencies. (a) $f_p = 0.5 \, GHz$, (b) $f_p = 1 \, GHz$, (c) $f_p = 3 \, GHz$, and (d) $f_p = 4 \, GHz$.}
	\label{fig:scan_beam}
\end{figure}

Leaky-wave holograms can be suitable choices for automotive sensors, coherent tomography, real-time spectrum analysis and tracking applications owing to their beam scanning capabilities, which  can be controlled by source frequency. Referring to the dispersion diagram in Fig. \ref{fig:dispersion_static}, for a pure-space modulated hologram, the corresponding phase constant increases by enhancing excitation frequency. This means that,  the radiation beam tends towards the end-fire direction at higher frequencies.
In temporally modulated holograms, beam scanning can be achieved through another mechanism. In this case, the beam direction varies by changing the pumping frequency ($f_p$) without the need to change the source frequency. 
The calculated dispersion diagrams for different values of $f_p$ are plotted in Fig. \ref{fig:dispersion_fp}. Observe that for a given phase constant, the (-1)-indexed curve tends towards the light-line as the pumping frequency increases. The dispersion curve for (+1)-indexed mode moves in opposite direction indicating that the incoming wave angle for reception case tends towards the broadside direction.
Figure \ref{fig:scan_beam} shows the calculated fields for different values of $f_p$. Results show that the beam direction changes from $\theta = 30^\circ$ to $\theta = 41^\circ$ with pumping frequency increasing from 0 GHz to 4 GHz. It should be noted that the source frequency is fixed at 18 GHz.

\section*{Conclusion}
In this paper, the generalized Floquet-wave expansion method is utilized to accurately calculate the propagation characteristics of temporally modulated holograms. 
The combination of holographic technique  as a synthesis method and Floquet-wave expansion can form a fully-analytical model for implementation of leaky-wave metasurfaces.
It has been shown that temporally modulated holograms can be effectively used as nonreciprocal antennas.  
This nonreciprocity can be obtained by properly displacing the dispersion curve so that the beam is directed at the desired spherical angle.
Owing to their embedded feeding networks, leaky-wave antennas can be appropriate alternatives to nonreciprocal metareflectors and transmitarrays  as radiators of integrated transceivers. 
\section*{Author contributions statement}

A. A. and H. O. developed the idea and theory, and wrote the manuscript. A. A. did the simulations. H. O. analyzed the results and supervised the project.

\section*{Additional information}
\textbf{Competing Interests:} The authors declare no competing interests.
%To include, in this order: \textbf{Accession codes} (where applicable); \textbf{Competing interests} (mandatory statement). 

%The corresponding author is responsible for submitting a \href{http://www.nature.com/srep/policies/index.html#competing}{competing interests statement} on behalf of all authors of the paper. This statement must be included in the submitted article file.

\bibliography{sample}

\newpage
%%%%%%%%%%%%%%%%%%%
%%%%%%%%%%%%%%%%%%%%%%%
%%%%%%%%%%%%%5%%%%%%%%%

\end{document}